\documentclass[10pt,final,doublecolumn]{IEEEtran}
\usepackage{makecell}
\usepackage{amsmath}
\usepackage{latexsym}
\usepackage{graphicx}
\usepackage{bbding}
\usepackage{indentfirst}
\usepackage{algorithm,algorithmic}
\usepackage{setspace}
\usepackage{float}
\usepackage{epstopdf}
\usepackage{amssymb}
\usepackage{amsfonts}
\usepackage{enumerate}
\usepackage{multicol}
\usepackage{color}
\usepackage{diagbox}
\usepackage{bm}
\usepackage{amssymb}
\usepackage{stfloats}
\usepackage{epstopdf}
\usepackage{threeparttable}
\usepackage{epstopdf}
\usepackage{threeparttable}
\usepackage{subfigure}
\usepackage{cite}

\IEEEoverridecommandlockouts
\allowdisplaybreaks[4]

\begin{document}
\title{Hybrid Near-Far Field 6D Movable Antenna Design Exploiting Directional Sparsity and Deep Learning}

\author{Xiaodan Shao, \IEEEmembership{Member,~IEEE}, Limei Hu, Yulong Sun, Xing Li, Yixiao Zhang, Jingze Ding, Xiaoming Shi, Feng Chen, Derrick Wing Kwan Ng,~\IEEEmembership{Fellow, IEEE}, Robert
	Schober,~\IEEEmembership{Fellow, IEEE}
	\vspace{-25pt}
	\thanks{X. Shao is with the Institute for Digital Communications, Friedrich-Alexander-University Erlangen-Nuremberg, 91054 Erlangen, Germany (email: shaoxiaodan@zju.edu.cn).}	
	
	\thanks{L.~Hu,  Y.~Sun, X. ~Li, and F. Chen are with College of Artificial Intelligence, Southwest University, Chongqing, 400715, China (emails: hlm0903@swu.edu.cn, sunyulong123@email.swu.edu.cn,  arbi2228445837@email.swu.edu.cn, fengchen.uestc@gmail.com). }
	
	\thanks{Y. Zhang is with the Department of Electrical and Computer Engineering, University of Waterloo, Waterloo, ON N2L 3G1, Canada (email: y3549zha@uwaterloo.ca).}	
	
	\thanks{J. Ding is with the School of Electronics, Peking University, Beijing 100871, China (e-mail: djz@stu.pku.edu.cn).}
	
	\thanks{X. Shi is with School of Science and Engineering, The Chinese University	of Hong Kong, Shenzhen, Guangdong 518172, China (e-mail: xiaomingshi@link.cuhk.edu.cn)}
	
	\thanks{D. W. K. Ng is with the School of Electrical Engineering and Telecommunications, University of New South Wales, Sydney, NSW 2052, Australia (e-mail: w.k.ng@unsw.edu.au).}
	
	\thanks{R. Schober is with the Institute for Digital Communications, Friedrich-Alexander-University Erlangen-Nurnberg (FAU), 91054
		Erlangen, Germany (email: robert.schober@fau.de).}
	}
	
	\maketitle
	
	\IEEEpeerreviewmaketitle
\begin{abstract}
		Six-dimensional movable antenna (6DMA) has been identified as a new disruptive technology for future wireless systems to support a large number of users with only a few antennas.
		However, the intricate relationships between the signal carrier wavelength and the transceiver region size lead to inaccuracies in traditional far-field 6DMA channel model, causing discrepancies between the model predictions and the hybrid-field channel characteristics in practical 6DMA systems, where users might be in the far-field region relative to the antennas on the same 6DMA surface, while simultaneously being in the near-field region relative to different 6DMA surfaces. Moreover, due to the high-dimensional channel and the coupled position and rotation constraints, the estimation of the 6DMA channel and the joint design of the 6DMA positions and rotations and the transmit beamforming at the base station (BS) incur extremely
		high computational complexity.
		To address these issues, we propose an
		efficient hybrid-field generalized 6DMA channel model, which accounts for planar-wave propagation within individual 6DMA surfaces and spherical-wave propagation among different 6DMA surfaces. Furthermore, by leveraging directional sparsity, we propose a low-overhead channel estimation algorithm that efficiently constructs a complete channel map for all potential antenna position-rotation pairs while limiting the training overhead incurred by antenna movement.		
		In addition, we propose a low-complexity design leveraging deep reinforcement learning (DRL), which facilitates the joint design of the 6DMA positions, rotations, and beamforming in a unified manner. Numerical
		results demonstrate the superiority of the proposed hybrid-field channel model, which achieves sum rates closely approaching that of the near-field channel model. The results also show that the proposed channel estimation algorithm can accurately recover the channel with lower computational complexity than traditional channel estimation algorithm. Moreover, the 6DMA system enhanced by the proposed DRL algorithm significantly outperforms existing flexible antenna systems, especially in the near-field region.
	\end{abstract}
	
	
		\section{Introduction}
	To accommodate the exponential growth in mobile traffic while ensuring ubiquitous connectivity, researchers in both
	industry and academic are dedicated to exploring revolutionary wireless network technologies. As we progress toward 6G communications, multiple-input multiple-output (MIMO) continues to play a vital role and spurs the development of various innovative technologies \cite{MIMOD, LA10, cyn,8949454,9325920}. For instance, by
	deploying more antennas than conventional massive MIMO within a compact space, extremely large-scale MIMO can achieve
	significantly high achievable rates. Passive MIMO, also known as intelligent reflecting surfaces (IRSs), can cost-effectively enhance network capacity \cite{shaos}.
	Moreover, cell-free massive MIMO strives to provide consistent coverage across several geographically dispersed access points \cite{cellfreeq}. However, these advancements are typically categorized into  fixed-position MIMO, which employs
	numerous fixed-position antennas (FPAs). While these technologies improve performance, they essentially come at the expense of increased hardware cost. Especially in fully-digital MIMO systems, each FPA, which is linked to a certain radio frequency (RF) transceiver consumes substantial power \cite{ret,ret1}. Moreover, fixed-position MIMO systems lack the flexibility to adapt antenna configurations
	based on the actual spatial distribution of user channels. As a result, FPAs struggle to satisfy the stringent efficiency and performance requirements of next-generation wireless networks.
	
	Recently, the six-dimensional movable antenna (6DMA) has been proposed as an innovative and practical solution to increase communication performance without the need for additional antennas. Specifically, this technology leverages the adaptability of antenna positions and orientations in three-dimensional (3D) space at transceivers to adaptively allocate antenna resources based on the actual spatial distribution of channels, thus providing a flexible and dynamic approach to managing the complexities of modern wireless communication environments \cite{shao20246d, 6dma_dis, rll}. Building on these advancements, the authors in \cite{rll} proposed a 6DMA-aided wireless sensing system, demonstrating that the 6DMA design significantly improves sensing performance. Subsequently, the authors in \cite{shaomaga, shim} further proposed a partially movable 6DMA solution aimed at striking an effective balance between performance enhancement and implementation cost. Then, the authors in \cite{jstsp} proposed a novel channel estimation method for 6DMA by leveraging the directional sparsity of 6DMA channels. Polarized 6DMA, proposed in \cite{shao2025polarized,shao2025polarized1}, enables polarforming to adaptively control the antenna’s polarization electrically and to tune its position and rotation mechanically.
	Meanwhile, the new research paradigm of 6DMA-aided unmanned aerial vehicles (UAVs), IRSs, and cell-free networks has recently been extensively studied \cite{chenzhi6, qingmma,lu2024wireless, shao2025tutorial,wang2025uav, 10891142, 10843383, luhai, zhang20246dma}. 
	Note that the 6DMA system can flexibly allocate antenna resources, fully leverage the antennas' directionality, array gain, and spatial multiplexing gain to achieve significant sensing/communication performance improvements compared to existing FPAs and movable and fluid antenna systems  \cite{10906511, 10709885,wu2024globally,hao1,qqw,weidong,10839251}.  
	
Despite their rich potential, existing 6DMA systems still encounter various practical challenges. For example, the far-field channel model, which relies on plane-wave assumptions, suffers from low accuracy due to its overly simplistic approach that captures only a small number of channel parameters \cite{nearfar, focus}. This model becomes particularly less accurate for 6DMA systems as the near-field region expands with increased dimensions of transmitter/receiver regions and higher carrier frequencies.  
However, applying the near-field channel model to 6DMA poses significant challenges and complexities \cite{nearfar1}. Since 6DMAs are capable of moving almost continuously in a given BS space and the available positions and rotations of 6DMAs are quite dense \cite{shao20246d, 6dma_dis}, modeling them leads to in an exponential increase in the number of parameters, which is proportional to the massive number of candidate positions and possible rotations\cite{dai,liuy}. In practice, a hybrid-field communication environment is
	more likely to be encountered in 6DMA systems, where users might be in the far-field region relative to antennas on one 6DMA surface, while being in the near-field region of other 6DMA surfaces. This variability arises because the positions and rotations of the 6DMA surfaces can be adjusted freely within a specified BS region. In other words, adopting models intended only for far‑field conditions causes a mismatch in the 6DMA channel. This mismatch highlights the pressing need for a highly integrated approach to model 6DMA channels as accurately as possible while maintaining low complexity across diverse operational environments.
	
On the other hand, to achieve optimal 6DMA performance, it is imperative to acquire accurate channel state information (CSI) linking every possible candidate antenna configuration with its user. Unlike FPAs that require estimating only a finite set of channel coefficients, 6DMA must acquire CSI over a continuous space with an enormous number of candidate antenna positions or rotations, which leads to high computational complexity and increased pilot overhead.
Moreover, based on the estimated CSI, the joint optimization of transmit beamforming, antenna position, and antenna rotation in 6DMA system remains largely unexplored. In practice, such a joint design presents a critical challenge for 6DMA applications due to the high-dimensional, non-convex nature of the problem, and the interdependency of position and rotation constraints. Fortunately, model-free deep learning has emerged as a powerful approach for addressing these complex design problems in wireless communications. Indeed, deep learning excels at automatically extracting meaningful features from large datasets and handling nonlinearities, thanks to its ability to embed specific biases into neural network architectures \cite{9432908,10555049}. In particular, deep reinforcement learning (DRL), a branch of deep learning, can significantly enhance the speed and performance of reinforcement learning algorithms \cite{rlwc}. When applied to wireless communication systems \cite{rlsv}, DRL offers distinct advantages:
1) Model-free Approach: Unlike traditional methods reliant on precise mathematical models, DRL adapts effectively to scenarios lacking accurate models or sufficient domain knowledge;
2) Dynamic Adaptability: DRL continuously adjusts its strategies to varying network conditions, ensuring consistently robust performance in dynamic environments;
3) Scalability: By leveraging deep neural networks' unified architecture, DRL can efficiently scale to accommodate different data characteristics and system environments, improving overall system scalability;
	Collectively, these attributes position DRL as a promising approach for optimizing communication systems, offering flexibility, adaptability, and scalability that traditional methods may struggle to achieve.
	
	In this paper, by considering a generalized 6DMA
	structure and leveraging the advantages of DRL, we address the intricate challenges associated with 6DMA channel modeling, channel estimation, and the joint design of transmit
	beamforming along with antennas' positions and rotations at the BS. The main contributions of this paper are summarized as follows.
	\begin{itemize}
		\item
		Firstly, since 6DMA surfaces can be freely positioned and rotated within the BS region, they may be spaced significantly apart from each other. This configuration results in users being positioned in the near-field region relative to  different 6DMA surfaces. On the other hand, since antennas on each 6DMA surface are positioned relatively close to one another, users typically remain in the far-field region relative these antennas.
		Therefore, we propose a hybrid-field 6DMA channel model which accounts for planar-wave propagation within individual  6DMA surfaces and spherical-wave propagation among different 6DMA surfaces. 
		This approach not only achieves greater accuracy than the far-field channel model, but also maintains lower complexity than the near-field channel model. 
		
		\item By exploiting the uneven channel power distribution properties through the concept of directional sparsity of 6DMA channels, we propose a channel estimation algorithm to construct a complete channel map for all potential antenna position-rotation pairs, enabling interference of the channel information between the user and any unmeasured position-rotation pairs in the hybrid-field 6DMA system.
		
		\item
		We introduce a novel optimization problem to maximize the sum rate of users by jointly optimizing transmit beamforming, 3D positions, and 3D rotations of 6DMA surfaces under antenna movement constraints. To address this problem, we propose a low-complexity DRL algorithm leveraging Advantage Actor-Critic (A2C) technology, which can iteratively maximize the sum rate guided by observed rewards.
		The proposed DRL-based approach learns directly from environmental feedback and offers efficient implementation without relying on explicit wireless environment models or specific mathematical formulations. This inherent flexibility allows it to scale effectively across various system settings.
		
		\item
		Finally, we evaluate the effectiveness of our proposed channel model, channel estimation, and DRL algorithm in optimizing the 6DMA positions, rotations, and transmit beamforming. Numerical experiments indicate that the achieved sum rate on the hybrid-field channel model closely approximates that of the ground-truth near-field model, but requiring fewer channel parameters. The results demonstrate that
		the proposed channel estimation algorithm can accurately obtain the channel.		
		Furthermore, our results illustrate that employing the DRL-enhanced 6DMA leads to significant improvements in the sum rate compared to traditional BS configurations employing FPAs or fluid antenna systems (FASs). 
	\end{itemize}
	
	The remainder of this paper is organized as follows. Section II presents the hybrid-field 6DMA channel model and signal model. The 6DMA channel estimation algorithm is proposed in Section III. In Section IV, we formulate an optimization problem for maximizing the sum rate through joint optimization of transmit beamforming and  positions/rotations of 6DMA surfaces, and introduce our proposed DRL-based algorithm to solve it. Section V presents numerical results and discussions. Finally, Section VI concludes the paper.
	
\emph{Notations}: The symbols $ \dagger$, \( (\cdot)^H \), and \( (\cdot)^T \) denote the conjugate, conjugate transpose, and transpose operations, respectively. The notation \( \mathbb{E}[\cdot] \) represents the expected value of a random variable, \( \left \| \cdot \right \| \) represents the Euclidean norm, \(|\cdot|\) represents the absolute value, \( \Re(\cdot) \) and \( \Im(\cdot) \) indicate the real and imaginary parts of a complex number, respectively, $\nabla$ denotes the gradient operation, $\frac{\partial (\cdot)}{\partial \mathbf{x}}$ denotes the partial derivative of a function with respect to vector $\mathbf{x}$, and $|\cdot|_{\mathrm{c}}$ denotes the cardinality of a set. 
	\section{System Model}
	\begin{figure}[t!]
		\centering
		\setlength{\abovecaptionskip}{0.cm}
		\includegraphics[width=3.6in]{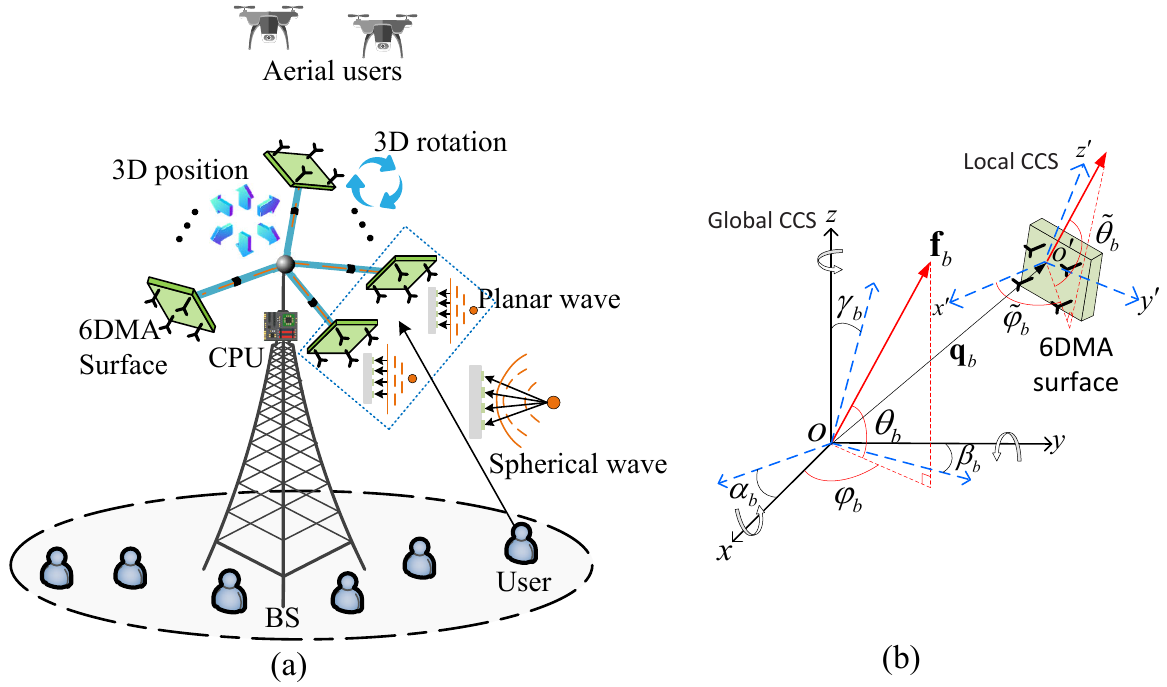}
		\caption{(a) A typical hybrid-field propagation environment in a 6DMA THz system; (b) The geometry of hybrid-field channel model.}
		\label{practical_scenario}
			\vspace{-0.49cm}
	\end{figure}
	In this section, we present the considered 6DMA system model along with the associated far-field, near-field, and hybrid-field channel models. Subsequently, we describe the signal model employed for channel estimation and data transmission.
	\vspace{-3 mm}
	\subsection{6DMA System Model}
	\vspace{-1 mm}
As shown in Fig. \ref{practical_scenario}(a), the 6DMA-enabled BS is composed of several 6DMA surfaces, denoted by \(\mathcal{B} = \{1, 2, \ldots, B\}\), that serve \(K\) single-antenna users in the downlink communication \cite{shao20246d, 6dma_dis}. Each 6DMA surface is configured as a uniform planar array (UPA) with \(N \geq 1\) antennas, which are represented by \(\mathcal{N} = \{1, 2, \ldots, N\}\). These surfaces are linked to a central processing unit (CPU) via extendable and rotatable rods, which enable flexible adjustments in their 3D positions and 3D rotations. 
	
The \(b\)‑th surface’s position and orientation are  
	\begin{align}\label{bb}
		\mathbf{q}_b=[x_b,y_b,z_b]^T\in\mathcal{C},~		\mathbf{u}_b=[\alpha_b,\beta_b,\gamma_b]^T, \forall b\in \mathcal{B},
	\end{align}
	where \(\mathcal{C}\) denotes the 3D antenna movement space. Here, \(x_b\), \(y_b\) and \(z_b\) are the \(b\)‑th surface’s center coordinates in the global Cartesian coordinate system (CCS) \( o\text{-}xyz \). The angles \( \alpha_b \), \( \beta_b \), and \( \gamma_b \) each range from 0 to \( 2\pi \). These angles represent rotation angles around the $x$-axis, $y$-axis, and $z$-axis, respectively (see Fig. \ref{practical_scenario}(a)).
	The rotation matrix \( \mathbf{R}(\mathbf{u}_b) \) with respect to \( \mathbf{u}_b \) is given by
\begin{align}\label{R}
	&\!\mathbf{R}(\mathbf{u}_b)\!=\!\begin{bmatrix}
		c_{\beta_b}c_{\gamma_b} & c_{\beta_b}s_{\gamma_b} & -s_{\beta_b} \\
		s_{\beta_b}s_{\alpha_b}c_{\gamma_b}-c_{\alpha_b}s_{\gamma_b} & s_{\beta_b}s_{\alpha_b}s_{\gamma_b}+c_{\alpha_b}c_{\gamma_b} & c_{\beta_b}s_{\alpha_b} \\
		c_{\alpha_b}s_{\beta_b}c_{\gamma_b}+s_{\alpha_b}s_{\gamma_b} & c_{\alpha_b}s_{\beta_b}s_{\gamma_b}-s_{\alpha_b}c_{\gamma_b} &c_{\alpha_b}c_{\beta_b} \\
	\end{bmatrix},\!\!
\end{align}
	where $c_x = \cos(x)$ and $s_x = \sin(x)$.
	
Let \(\bar{\mathbf{r}}_{n}\) denote the position of the \(n\)-th antenna on the \(b\)-th 6DMA surface in its local CCS \(o'\text{-}x'y'z'\). The  position of the \( n \)-th antenna on the \( b \)-th 6DMA surface in the global CCS is then expressed as
	\begin{align}\label{nwq}
		\mathbf{r}_{b,n}(\mathbf{q}_b,\mathbf{u}_b)=\mathbf{q}_b+\mathbf{R}
		(\mathbf{u}_b)\bar{\mathbf{r}}_{n},~n\in\mathcal{N},~b \in\mathcal{B}.
	\end{align}

	\subsection{Channel Model}
Each user is assumed to employ a single FPA and a line-of-sight (LoS) channel is present between the user's location and the 6DMA-enabled BS. We begin by
presenting the far-field and near-field 6DMA channel models,
and subsequently introduce the proposed hybrid-field channel
model.
	\subsubsection{Far-Field 6DMA Channel Model}
	Let $\phi\in[-\pi,\pi]$ and $\theta\in[-\pi/2,\pi/2]$ represent the azimuth and elevation angles, respectively, of a signal originating from any user arriving with regard to BS center $o$. The unit-length direction-of-arrival (DOA) vector corresponding to the direction $(\theta, \phi)$ is given by 
	\begin{align}\label{KM}
		\hat{\mathbf{f}}=[\cos(\theta)\cos(\phi), \cos(\theta)\sin(\phi), \sin(\theta)]^T.
	\end{align}
Then, the steering vector for the \( b \)-th 6DMA surface can be expressed as
	\begin{align}\label{gen}
		\hat{\mathbf{a}}(\mathbf{q}_b,\mathbf{u}_b)=& \left[e^{-j\frac{2\pi}{\lambda}
			\hat{\mathbf{f}}^T\mathbf{r}_{b,1}(\mathbf{q}_b,\mathbf{u}_b)},
		\cdots, e^{-j\frac{2\pi}{\lambda}\hat{\mathbf{f}}^T
			\mathbf{r}_{b,N}(\mathbf{q}_b,\mathbf{u}_b)}\right]^T,
	\end{align}
	where $\lambda$ denotes the carrier wavelength. Then, the far-field channel between any user's location and the 6DMA-enabled BS is expressed as \cite{shao20246d}
	\begin{align}\label{uk}
		&\!\!\mathbf{h}^{\mathrm{far}}(\mathbf{q},\mathbf{u})\!
		=\nonumber\\
		&\!\!\nu
		e^{-j\frac{2\pi d}{\lambda}}\!\left[\!\sqrt{\hat{g}(\mathbf{u}_1)}
		\hat{\mathbf{a}}^T(\mathbf{q}_1,\mathbf{u}_1),
		\!\cdots\!,\sqrt{\hat{g}(\mathbf{u}_B)}\hat{\mathbf{a}}^T(\mathbf{q}_B,\mathbf{u}_B)
		\!\right]^T,\!\!
	\end{align}
	with $
		{\mathbf{q}}=[\mathbf{q}_1^T,\cdots,\mathbf{q}_B
		^T]^T\in \mathbb{R}^{3B\times 1}$ and $
		\mathbf{u}=[\mathbf{u}_1^T,\cdots,\mathbf{u}_B^T]^T\in \mathbb{R}^{3B\times 1}$.
	In the above, $d$ denotes the distance between any user's location and the reference position of the BS, $\nu$ denotes the path gain, and $\hat{g}(\mathbf{u}_b)$ denote the effective antenna gain for the $b$-th 6DMA surface in linear scale. To define $\hat{g}(\mathbf{u}_b)$, we project the DOA vector, \( \hat{\mathbf{f}} \), onto the local CCS of the \( b \)-th 6DMA surface as
	\begin{align}
		\hat{\mathbf{f}}_b=\mathbf{R}^{-1}(\mathbf{u}_b)
		\hat{\mathbf{f}}.
	\end{align}
	Then, we represent $\hat{\mathbf{f}}$ in the spherical coordinate system as
	\begin{align}\label{KM3}
		\hat{\mathbf{f}}_b=[\cos(\hat{\theta}_{b})\cos(\hat{\phi}_{b}), \cos(\hat{\theta}_{b})\sin(\hat{\phi}_{b}), \sin(\hat{\theta}_{b})]^T,
	\end{align}
	where $\hat{\theta}_{b}$ and $\hat{\phi}_{b}$ represent the corresponding DOAs in the local CCS. Next, $\hat{g}(\mathbf{u}_b)$ can be represented as 
	\begin{align}\label{gm}
		\hat{g}(\mathbf{u}_b)=10^{\frac{A(\hat{\theta}_{b}, \hat{\phi}_{b})}{10}},
	\end{align}
	where \( A(\hat{\theta}_{b}, \hat{\phi}_{b}) \) denotes the effective gain of each antenna on the \( b \)-th 6DMA surface, measured in decibels isotropic (dBi) \cite{shao20246d}. 
		
	%In the above, $\nu=(\frac{\lambda}{4\pi d_0})^2(\frac{d_0}{\mathbf{v}})^\epsilon$ is the path gain, where $\mathbf{v}$ denotes the location of user in global CCS, and $\varsigma$ denotes the path loss exponent. When $\varsigma=2$, $\sqrt{\nu}$ reduces to  $\sqrt{\nu}=\frac{\lambda}{4\pi \mathbf{v}}$.
	
	\subsubsection{Near-Field 6DMA Channel Model}
	The spherical-wave model, which calculates the channel responses for all antenna pairs between the transmitter and receiver, is commonly employed for near-field communication due to its accuracy at shorter distances \cite{focus}.
	In the following, we consider the near-field wireless channel model.
	
	In the near-field 6DMA channel model, we define $\phi_{b,n}\in[-\pi,\pi]$ and $\theta_{b,n}\in[-\pi/2,\pi/2]$ as the azimuth and elevation angles, respectively, of a signal from any user arriving at the BS with respect to the $n$-th antenna on the $b$-th 6DMA surface. Then, the near-field channel model is given by
	\begin{align}
		& \mathbf{h}^{\mathrm{near}}(\mathbf{q},\mathbf{u})	= [\nu_{1,1}{\sqrt{\bar{g}(\mathbf{q}_1,\mathbf{u}_1)}}
		^{-j\frac{2\pi}{\lambda}
			\bar{d}_{1,1}},
		\cdots, \nu_{b,n}{\sqrt{\bar{g}(\mathbf{q}_b,\mathbf{u}_b)}}\nonumber\\
		&
		e^{-j\frac{2\pi}{\lambda}
			\bar{d}_{b,n}},\cdots,	\nu_{B,N}{\sqrt{\bar{g}(\mathbf{q}_B,\mathbf{u}_B)}}
		e^{-j\frac{2\pi}{\lambda}
			\bar{d}_{B,N}}]^T, \label{hnn}
	\end{align}
where \( \bar{d}_{b,n} \) represents the distance between any user and the \(n\)-th antenna on the \(b\)-th 6DMA surface at the BS, $\nu_{b,n}$ denotes the corresponding path gain, and the antenna gain is given by \( \bar{g}(\mathbf{q}_b, \mathbf{u}_b) = 10^{\frac{A(\bar{\theta}_{b,n}, \bar{\phi}_{b,n})}{10}} \).  Similar to \eqref{gm}, the angles \( \bar{\theta}_{b,n} \) and \( \bar{\phi}_{b,n} \) can be obtained by projecting DOA vector \( \bar{\mathbf{f}}_{b,n}\!=\![\cos(\theta_{b,n})\cos(\phi_{b,n}), \cos(\theta_{b,n})\sin(\phi_{b,n}),\sin(\theta_{b,n})]^T \) onto the local CCS of the \( b \)-th 6DMA surface through $\mathbf{R}(\mathbf{u}_b)^T\bar{\mathbf{f}}_{b,n}$.

\subsubsection{Hybrid-Field 6DMA Channel Model}
	In the 6DMA system, the far-field  model is generally
	inaccurate due to its simplistic assumptions, while the near field model imposes high modeling complexity. Moreover, since 6DMAs can move almost continuously in a
	high-dimensional BS
	space, the users might be in the far-field region relative to antennas on the same 6DMA surface, while being in the near-field region of different 6DMA surfaces. Thus, in the following, we propose the hybrid-field wireless channel model using fewer parameters, which accounts for far-field channel model within one 6DMA surface and near-field channel model among 6DMA surfaces by assigning a unique signal transmission angle to every surface.
	
%	The main idea is to assign distinct signal transmission angles to each 6DMA surface, which directly determines its antenna gain and channel phase shift. In contrast, since the centers of all 6DMA surfaces lie within a region that is negligible compared with the distance between the user and the BS, the corresponding large-scale path losses vary by an imperceptible margin. Therefore, we model a common path gain for all 6DMA surfaces to reduce the number of channel parameters. 
	Specifically, in the proposed hybrid-field 6DMA channel model, let $\phi_{b}\in[-\pi,\pi]$ and $\theta_{b}\in[-\pi/2,\pi/2]$ denote the azimuth and elevation angles, respectively, of a signal from any user arriving at the BS with respect to the $b$-th 6DMA surface (see Fig. \ref{practical_scenario}(b)).  The unit-length DOA vector corresponding to the direction $(\theta_b, \phi_b)$ is given by
	\begin{align}
	{\mathbf{f}}_{b}=[\cos(\theta_{b})\cos(\phi_{b}), \cos(\theta_{b})\sin(\phi_{b}), \sin(\theta_{b})]^T.\label{KMh}
    \end{align}
Consequently, the steering vector of the \( b \)-th 6DMA surface is given by
		\begin{align}
			\!\!\mathbf{a}(\mathbf{q}_b,\!\mathbf{u}_b)\!= \!\left[\!e^{-j\frac{2\pi}{\lambda}
				{\mathbf{f}}_{b}^T\mathbf{r}_{b,1}(\mathbf{q}_b,\mathbf{u}_b)},
			\!\cdots\!, e^{-j\frac{2\pi}{\lambda}{\mathbf{f}}_{b}^T
				\mathbf{r}_{b,N}(\mathbf{q}_b,\mathbf{u}_b)}
			\!\right]^T. \!\!\!\label{genhtt}
		\end{align}
	
Based on \eqref{KMh} and \eqref{genhtt}, we construct the following hybrid-field channel model between any user's location and the BS:
	\begin{align}
		&\!\!\!\!\!\!\!\mathbf{h}(\mathbf{q},\mathbf{u})=
		\nu[{ \sqrt{g(\mathbf{q}_1,\mathbf{u}_1)}e^{-j\frac{2\pi}{\lambda}
				d_{1}}}
		\mathbf{a}^T(\mathbf{q}_1,
		\mathbf{u}_1),\cdots,\nonumber\\
		&{ \sqrt{g(\mathbf{q}_B,\mathbf{u}_B)}e^{-j\frac{2\pi}{\lambda}
				d_{B}}}
		\mathbf{a}^T(\mathbf{q}_B,\mathbf{u}_B)
		]^T\in\mathbb{C}^{NB\times 1},\label{ap}
	\end{align}
where $d_{b}$ denotes the distance between any user's location and the $b$-th 6DMA surface, and the corresponding antenna gain is given by \( g(\mathbf{q}_b,\mathbf{u}_b)=10^{\frac{A(\tilde{\theta}_{b}, \tilde{\phi}_{b})}{10}} \).  Similar to \eqref{gm}, the angles \( \tilde{\theta}_{b} \) and \( \tilde{\phi}_{b} \) are obtained by projecting DOA vector \( {\mathbf{f}}_{b}\) in \eqref{KMh} onto the local CCS of the \( b \)-th 6DMA surface adopting $
\mathbf{R}
(\mathbf{u}_b)^T{\mathbf{f}}_{b}$ (see Fig. \ref{practical_scenario}(b)).
	
In practice, the spatial multiplexing capability of a 6DMA system is more accurately characterized by adopting the hybrid-field 6DMA channel model compared with the far-field 6DMA channel model. This improvement is due to the consideration of different transmission angles to each 6DMA surface, which results in signals that are more spatially separated at the 6DMA receiver. Consequently, this spatial separation facilitates a well-conditioned channel that maximizes spatial multiplexing gain. In contrast, the far-field 6DMA channel model assumes identical transmission angles to all 6DMA surfaces, and when users are in close proximity, the rank of the multiuser channel is often less than $\min\{K,NB\}$ with $K$ being the number of users.
	
	{{\textbf{Remark 1}}}:
	When $B=1$, the proposed hybrid-field 6DMA channel model reduces to the conventional far-field 6DMA model in \cite{shao20246d}; when $N=1$, the hybrid-field model effectively reduces to the near-field 6DMA channel model. 
	\subsection{Signal Model}
	\subsubsection{Signal Model for Channel Estimation}
	Before downlink data transmission, users first transmit orthogonal training pilots to the BS over $T$ time slots for CSI acquisition. Therefore, channel estimation
	can be performed independently from each user and we can consider an arbitrary user without loss of generality \cite{9117124}. During this phase, all 6DMA surfaces traverse a small number of
	$M$ different position-rotation pairs, denoted by $\{\mathbf{q}_m,\mathbf{u}_m\}_{m=1}^M$, to collect the corresponding channel
	measurements for reconstructing the channel information for all available antenna locations and rotations.
	
	For the \(m\)-th 6DMA position-rotation pair, the received signal is given by
	\begin{align}\label{6}
		\mathbf{y}_m = \mathbf{h}_m(\mathbf{q}_m,\mathbf{u}_m)x+ \mathbf{z}_m,
	\end{align}
	where
	\begin{align}
		\mathbf{h}_m(\mathbf{q}_m,\mathbf{u}_m)=\nu{ \sqrt{g(\mathbf{q}_m,\mathbf{u}_m)}e^{-j\frac{2\pi}{\lambda}
				d_{m}}}
		\mathbf{a}^T(\mathbf{q}_m,
		\mathbf{u}_m),
	\end{align}
	denotes the channel between a user's location and the $m$-th 6DMA surface. In \eqref{6}, $x\in\mathbb{C}^1$ is the transmitted pilot and $\mathbf{z}_m$ denotes an additive white Gaussian noise (AWGN) vector whose elements have zero mean and variance $\sigma^2$. Based on \eqref{6}, the 6DMA channel of the system can be reconstructed by processing \(\mathbf{y}_m\) for \(m \in \mathcal{M}=\{1,2,\cdots,M\}\).
	\subsubsection{Signal Model for Data Transmission}
	As shown in Fig. \ref{practical_scenario}(a), we consider a multi-user downlink communication system adopting the hybrid-field 6DMA channel model. The signal transmitted by the 6DMA-BS can be written as
	\begin{align}\label{X}
		\bar{\mathbf{x}}=\sum_{k=1}^K\mathbf{w}_k\bar{x}_k,
	\end{align}
	where $\bar{x}_k\in \mathbb{C}^{1}$ is the normalized data symbol for the $k$-th
	user, i.e., $\mathbb{E}[|\bar{x}_k^2|]=1$, and $\mathbf{w}_k\in\mathbb{C}^{NB\times 1}$ is the precoding (i.e., beamforming) vector for $\bar{x}_k$, and $\mathbb{E}[\bar{x}_j^{ \dagger}\bar{x}_k]=0,
	\forall j\neq k$.
	
	According to the channel input \(\bar{\mathbf{x}}\) in \eqref{X}, the received signal at the $k$-th user with the hybrid-field 6DMA channel model is given by
	\begin{align}\label{X1}
		y_k=
		\mathbf{h}_k^H(\mathbf{q},\mathbf{u})\sum_{j=1}^K
		\mathbf{w}_j\bar{x}_j+n_k,~k\in \mathcal{K},
	\end{align}
	where $\mathcal{K}=\{1,2,\cdots,K\}$, $n_k$ is the AWGN with zero mean and variance $\sigma^2$, and $\mathbf{h}_k^H(\mathbf{q},\mathbf{u})$ is the channel between the \(k\)-th user and the BS, generated from \eqref{ap}. Accordingly, the achievable rate of the $k$-th user is given by
	\begin{align}\label{r}
		&C_k(\mathbf{w}_k,\mathbf{q},\mathbf{u})= \log_2\left(1+
		\frac{|({\mathbf{h}_k}(\mathbf{q},\mathbf{u})^H
			\mathbf{w}_k|^2}{\sum_{j\neq k}|({\mathbf{h}_k}(\mathbf{q},\mathbf{u})^H
			\mathbf{w}_j|^2+\sigma^2}\right).
	\end{align}

	\vspace{-3 mm}
\section{6DMA Channel Estimation}
To achieve favorable 6DMA positions and rotations, acquiring the channel information of all available antenna locations and rotations is essential. However, the number of available measurements for position-rotation pairs is typically limited. Consequently, it is imperative to construct a comprehensive channel map covering all potential antenna position-rotation configurations, thereby enhance accurate inference of the channel characteristics for unmeasured pairs.

To facilitate this reconstruction, for each \(b \in \mathcal{B}\), parameters \(d_b\) and \((\phi_b, \theta_b)\) in \eqref{ap} can be derived from \(d\) and \((\phi, \theta)\), respectively, according to the channel's geometric relationships. Therefore, the hybrid-field 6DMA channel becomes a function of path gain \(\nu\), user-BS distance \(d\), and the signal's azimuth and elevation angles, \(\phi\) and \(\theta\), relative to the BS center. Therefore, by accurately determining \(\nu\), \(d\), \(\phi\), and \(\theta\) (i.e., the users' locations) from the received signal, the complete 6DMA channel between users and all possible 6DMA positions and rotations at the BS can be reconstructed. 

Building on this foundation, the proposed scheme leverages the directional sparsity property (will be discussed in Section III-A) of 6DMA channels and is implemented in two steps. First, each 6DMA position‑rotation pair (i.e., its surface) produces an independent coarse estimate, which offers scalability and low computational complexity by treating each 6DMA position-rotation pair independently. Then, a refined parameter estimation is carried out by adopting those 6DMA position-rotation pairs that exhibit strong received signals. This collaborative approach improves channel reconstruction accuracy by exploiting the uneven channel power distribution resulting from directional sparsity. In the following, we introduce the details of the designed channel estimation algorithm.
\subsection{Unequal Channel Power from Directional Sparsity}
The 6DMA  system exhibits an uneven channel power distribution across its surfaces due to  inherent directional sparsity (see Fig. \ref{heatmap} in Section V). Specifically, owing to the antenna directivity in 6DMA \cite{jstsp}, each user is typically served by only a limited subset of the available 6DMA position-rotation pairs, denoted by the set $\mathcal{W}\subseteq \mathcal{M}$
(i.e., antenna gain $g(\mathbf{q}_m,\mathbf{u}_m)\neq 0$ for all $ m\in \mathcal{W}$). In contrast, for the remaining candidate position-rotation pairs, which may either face in the opposite direction of the user or be blocked by obstacles towards the user, the effective channel gains of the user are significantly weaker and thus can be ignored (i.e.,  $g(\mathbf{q}_m,\mathbf{u}_m)=0$ for all $m\in \mathcal{W}^{\mathrm{c}}$, where $\mathcal{W} \cup \mathcal{W}^{\mathrm{c}} = \mathcal{M}$ and $ \mathcal{W} \cap \mathcal{W}^{\mathrm{c}} = \emptyset
$). We refer to this characteristic as {\textit{directional sparsity}}, which is unique to 6DMA systems \cite{shao20246d}. For example, in Fig. \ref{practical_scenario}(a), a user located on the ground can establish a strong channel with the 6DMA surface directed toward it, whereas the channel with the 6DMA surface oriented skyward is much weaker and can be approximated as zero.

Such directional sparsity causes negligible signal power to be observed at some 6DMA position-rotation pairs for a given user. This uneven power distribution, along with the high dimensionality of candidate antenna position-rotation pairs, significantly impacts both the performance and computational complexity of 6DMA channel estimation. Therefore, an alternative channel estimation approach is designed in the following for 6DMA  systems, which exploits directional sparsity and the resulting spatial non-stationarities to achieve an optimal trade-off between performance and computational complexity, even when a large number of candidate antenna positions and rotations are involved.
\subsection{6DMA Surface-Wise Estimation}
Since the antennas on the same 6DMA surface experience similar channel gains, we treat each surface individually and localize the visible users on each surface employing a maximum likelihood (ML) algorithm \cite{mll}. Specifically, 
based on \eqref{6}, we reformulate the 6DMA channel estimation task as a parameter estimation problem. At the $m$-th 6DMA position-rotation pair, the ML estimate of \((d, \phi, \theta)\), denoted by $(\hat{d}^{(m)}, \hat{\phi}^{(m)},\hat{\theta}^{(m)})$, is given by
\begin{align}\label{mo3}
	&\!\!\!(\hat{d}^{(m)}, \hat{\phi}^{(m)},\hat{\theta}^{(m)})=\arg\min_{d,\phi,\theta} \nonumber\\
	& \| \mathbf{y}_m\! -\! \nu x{ \sqrt{g(\mathbf{q}_m,\mathbf{u}_m)}e^{-j\frac{2\pi}{\lambda}
			d_{m}}}\mathbf{a}(\mathbf{q}_m,
	\mathbf{u}_m) \|^2.
\end{align}
To facilitate the search, we introduce a coarse search grid \(\Xi_{\mathrm{H}}\) defined as follows:
\begin{align}
	\Xi_{\mathrm{H}} = \{(d,\phi,\theta)| d &= D_{\min}, D_{\min} + \Delta d_{\mathrm{H}}, \ldots, D_{\max}; \nonumber\\
	\phi& = -\pi, -\pi + \Delta \phi_{\mathrm{H}}, \ldots, \pi;\nonumber\\
	\theta& = -\frac{\pi}{2}, -\frac{\pi}{2} + \Delta \theta_{\mathrm{H}}, \ldots, \frac{\pi}{2} \},
\end{align}
where $\Delta d_{\mathrm{H}}$, $\Delta \phi_{\mathrm{H}}$, and $\Delta \theta_{\mathrm{H}}$ are the step sizes for the distance and angular dimensions of the grid $\Xi_{\mathrm{H}}$, respectively, and the interval \([D_{\min}, D_{\max}]\) constrains the user's possible distance range. 

From \eqref{mo3}, we note that the objective function is a quadratic polynomial in \(\nu\). We therefore derive its closed-form solution. Specifically, taking the derivative of \eqref{mo3} with respect to $\nu$ and equating it to zero, we obtain
\begin{align} \label{mo5}
	\hat{\nu} = \frac{
		(x{ \sqrt{g(\mathbf{q}_m,\mathbf{u}_m)}e^{-j\frac{2\pi}{\lambda}
				d_{m}}} \mathbf{a}(\mathbf{q}_m, \mathbf{u}_m))^H \mathbf{y}_m}{
		\| x{ \sqrt{g(\mathbf{q}_m,\mathbf{u}_m)}e^{-j\frac{2\pi}{\lambda}
				d_{m}}} \mathbf{a}(\mathbf{q}_m, \mathbf{u}_m) \|^2}.
\end{align}
By substituting \eqref{mo5} into \eqref{mo3}, the optimization problem in \eqref{mo3} can be reformulated as
\begin{align}\label{hu8}
	&(\hat{d}^{(m)}, \hat{\phi}^{(m)}, \hat{\theta}^{(m)}) = \arg\max_{(d,\phi,\theta) \in \Xi_{\mathrm{H}}} \nonumber\\
	& \frac{| (x \sqrt{g(\mathbf{q}_m,\mathbf{u}_m)} e^{-j\frac{2\pi}{\lambda}d_{m}} \mathbf{a}(\mathbf{q}_m, \mathbf{u}_m))^H \mathbf{y}_m |^2}{\| x \sqrt{g(\mathbf{q}_m,\mathbf{u}_m)} e^{-j\frac{2\pi}{\lambda}d_{m}} \mathbf{a}(\mathbf{q}_m, \mathbf{u}_m) \|^2}.
\end{align}

\subsection{Multi-Surface Estimation Refinement}
After performing surface-wise estimation, the surfaces then collaborate to further refine channel estimation, thereby enhancing estimation accuracy. However, due to directional sparsity, the channel estimation derived from certain 6DMA position-rotation pairs that receive insufficient  (or any) power from the user may be inaccurate. To address this problem, a distance-based clustering method in \cite{diss} is adopted to obtain the final refined estimates of \((d, \phi, \theta, \nu)\), denoted by \((\bar{d}, \bar{\phi}, \bar{\theta}, \bar{\nu})\). Specifically, once the set \(\{\hat{d}^{(m)}, \hat{\phi}^{(m)}, \hat{\theta}^{(m)}\}_{m=1}^M\) is obtained via \eqref{hu8}, each triple \((\hat{d}^{(m)}, \hat{\phi}^{(m)}, \hat{\theta}^{(m)})\) is converted into its corresponding Cartesian coordinate, yielding 
\begin{align}
	\{\hat{\mathbf{v}}_m=[\hat{x}_m, \hat{y}_m, \hat{z}_m]^T\}_{m=1}^M,
\end{align}
where \(\hat{\mathbf{v}}_m\) denotes the estimated user position  derived from the \(m\)‑th 6DMA position‑rotation pair.
Then, a single cluster \(\mathcal{S}_1\) is first initialized by exploiting the estimate \(\hat{\mathbf{v}}_1\), with its center defined as \(\mathbf{c}_1 = \hat{\mathbf{v}}_1\). For each subsequent estimate \(\hat{\mathbf{v}}_m\), the Euclidean distance to all existing cluster centroids \(\mathbf{c}_i\), \(i = 1, \dots, N_c\), is computed as follows
\begin{align}
	D^{(m,i)} = \|\hat{\mathbf{v}}_m - \mathbf{c}_i\|,
\end{align}
where \(N_c\) denotes the number of clusters formed up to the current iteration.
The index corresponding to the minimum distance is given by
\begin{align}\label{minn}
	i^* = \arg\min_{1 \le i \le N_c} D^{(m,i)}.
\end{align}
If the minimum distance \(D_{\min}^{(m,i^*)}\) satisfies \(D_{\min}^{(m,i^*)} \le \epsilon\) for a predefined threshold \(\epsilon>0\), the estimate \(\hat{\mathbf{v}}_m\) is added to cluster \(\mathcal{S}_{i^*}\) and the corresponding cluster center is updated as
\begin{align}\label{vv0}
	\mathbf{c}_{i^*} \leftarrow \frac{1}{\lvert \mathcal{S}_{i^*}\rvert_\mathrm{c}}\sum_{i \in \mathcal{S}_{i^*}} \hat{\mathbf{v}}_i.
\end{align}
Conversely, if \(D_{\min}^{(m,i^*)} > \epsilon\), \(\hat{\mathbf{v}}_m\) is considered too distant from all existing clusters, prompting the creation of a new cluster \(\mathcal{S}_{N_c+1}=\{m\}\) with \(\mathbf{c}_{N_c+1} =\hat{\mathbf{v}}_m\). After processing estimates from all 6DMA position-rotation pairs, the cluster index with the largest cardinality is identified as
\begin{align}\label{mmax}
	m_{\max} = \arg\max_{1 \le i \le N_c} \lvert \mathcal{S}_i\rvert_\mathrm{c}.
\end{align}
Consequently, the cluster that contains the largest number of 6DMA position-rotation pairs is denoted by \(\mathcal{S}_{m_{\max}}\), where \(\lvert \mathcal{S}_{m_{\max}}\rvert_\mathrm{c} = F\). 
Since reliable estimates of the true user channel tend to concentrate within a single dense cluster, whereas inaccurate or noisy estimates are scattered among smaller clusters, 
the center of cluster $\mathcal{S}_{m_{\max}}$, denoted by \(\mathbf{c}_{m_{\max}}\), is designated as the final user coordinate estimate \((\hat{x},\hat{y},\hat{z})\) (i.e., $(\hat{d}, \hat{\phi}, \hat{\theta})$ in polar coordinates). 
\begin{algorithm}[t]
	\caption{Directional-Sparsity-Driven 6DMA Channel Estimation Algorithm}
	\label{alg:6dma_estimation}
	\begin{algorithmic}[1]
		\STATE \textbf{Input:} $\Xi_{\mathrm{H}}$, $\Xi_{\mathrm{L}}$, $\epsilon$, $M$
		\STATE \textbf{Initialization:}  $\mathcal{S}_{1}=1$, $N_c=1$
		\STATE \emph{Step 1: 6DMA Surface-Wise Estimation}
		\STATE All 6DMA surfaces traverse \( M \) position-rotation pairs to collect channel measurements.
		\FOR{$m = 1$ to $M$}
		\STATE Obtain \((\hat{d}^{(m)}, \hat{\phi}^{(m)}, \hat{\theta}^{(m)})\) from \eqref{hu8} and \(\hat{\nu}\) from \eqref{mo5}.
		\ENDFOR
		\STATE \emph{Step 2: Multi-Surface Estimation Refinement}
		\STATE $\mathbf{c}_1=\hat{\mathbf{v}}_1$.
		\FOR{$m = 2$ to $M$}
		\STATE Determine the index $i^*$ according to \eqref{minn}.
		\IF{$D_{\min}^{(m,i^\star)} \leq \varepsilon$}
		\STATE $\mathcal{S}_{i^\star} = \mathcal{S}_{i^\star} \cup m$, and update $\mathbf{c}_{i^*}$ according to \eqref{vv0}.
		\ELSE
		\STATE $N_c=N_c+1$, $\mathcal{S}_{N_c} = \{m\}$ and \(\mathbf{c}_{N_c} =\hat{\mathbf{v}}_m\).
		\ENDIF
		\ENDFOR
		\STATE Identify $\mathcal{S}_{m_{\max}}$ according to \eqref{mmax}.
		\STATE Calculate $\bar{d}$, $\bar{\phi}$, $\bar{\theta}$, $\hat{\nu}$ via \eqref{mo1} and \eqref{mo12}.
		\STATE Reconstruct \(\hat{\mathbf{h}}$ at any position-rotation pair \((\mathbf{q}, \mathbf{u})\) via \eqref{go9}.
		\STATE \textbf{Return} $\bar{d}$, $\bar{\phi}$, $\bar{\theta}$, $\hat{\nu}$, and $\hat{\mathbf{h}}(\mathbf{q}, \mathbf{u})$.
	\end{algorithmic}
\end{algorithm}

Once coarse estimate $(\hat{d}, \hat{\phi}, \hat{\theta})$ is obtained, more accurate values, $(\bar{d}, \bar{\phi}, \bar{\theta})$, can be derived via multiple-surface-based joint channel estimation with smaller step sizes, i.e., $\Delta d_{\mathrm{L}}<\Delta d_{\mathrm{H}}$, $\Delta \phi_{\mathrm{L}}<\Delta \phi_{\mathrm{H}}$, and $\Delta \theta_{\mathrm{L}}<\Delta \theta_{\mathrm{H}}$. The corresponding fine search grid \(\Xi_{\mathrm{L}}\) is given by 
\begin{align}
	&\Xi_{\mathrm{L}} = \{(d,\phi,\theta)| \nonumber\\
	&d= \hat{d} - \overline{D}, \ldots, \hat{d} - \Delta d_{\mathrm{L}}, \hat{d}, \hat{d} + \Delta d_{\mathrm{L}}, \ldots, \hat{d} + \overline{D}; \nonumber\\
	&\phi = \hat{\phi} - \overline{\Phi}, \ldots, \hat{\phi} - \Delta \phi_{\mathrm{L}}, \hat{\phi}, \hat{\phi} + \Delta \phi_{\mathrm{L}}, \ldots, \hat{\phi} + \overline{\Phi} ;\nonumber\\
	&\theta = \hat{\theta} - \overline{\Theta}, \ldots, \hat{\theta} - \Delta \theta_{\mathrm{L}}, \hat{\theta}, \hat{\theta} + \Delta \theta_{\mathrm{L}}, \ldots, \hat{\theta} + \overline{\Theta} \},
\end{align}
where \(2\overline{D}\), \(2\overline{\Phi}\), and \(2\overline{\Theta}\) denote the ranges for distance and angles, respectively, of the grid \(\Xi_{\mathrm{L}}\). Based on \(\Xi_{\mathrm{L}}\), the final parameter estimates are jointly determined by multiple 6DMA surfaces as follows:
\begin{subequations}
	\label{moww}
	\begin{align}\label{mo1}
		(\bar{d}, \bar{\phi}, \bar{\theta}) &= \arg\max_{(d,\phi,\theta)\in \Xi_{\mathrm{L}}} \frac{\left| ( \mathbf{a}_{\mathrm{L}}(\mathbf{q},\mathbf{u}))^H \mathbf{y}_{\mathrm{L}} \right|^2}{\left\|  \mathbf{a}_{\mathrm{L}}(\mathbf{q},\mathbf{u}) \right\|^2},\\
		\bar{\nu}&= \frac{ ( \mathbf{a}_{\mathrm{L}}(\mathbf{q},\mathbf{u}))^H \mathbf{y}_{\mathrm{L}} }{\left\|  \mathbf{a}_{\mathrm{L}}(\mathbf{q},\mathbf{u}) \right\|^2},\label{mo12}
	\end{align}
\end{subequations}
where $\mathbf{y}_{\mathrm{L}}\! =\! [\mathbf{y}_{j_1}^T, \ldots, \mathbf{y}_{j_F}^T]^T$, and $\mathbf{a}_{\mathrm{L}} = [\mathbf{a}_{j_1}^T{ \sqrt{g(\mathbf{q}_{j_1},\mathbf{u}_{j_1})}e^{-j\frac{2\pi}{\lambda}
		d_{j_1}}}x,  \ldots, 
\mathbf{a}_{j_F}^T \sqrt{g(\mathbf{q}_{j_F},\mathbf{u}_{j_F})}e^{-j\frac{2\pi}{\lambda}
	d_{j_F}}x ]^T$ denote the combination of received signals and steering vectors of 6DMA position-rotation pairs in set $\mathcal{S}_{\max}$, respectively. Here, \(j_1, j_2, \dots, j_F\) are the indices of the 6DMA position-rotation pairs that belong to the cluster \(\mathcal{S}_{m_{\max}}\). 

Given $\bar{d}$, $\bar{\phi}$, $\bar{\theta}$, and $\bar{\nu}$, the channel between the user and a 6DMA surface at an arbitrary position-rotation pair \((\mathbf{q}, \mathbf{u})\) can be reconstructed according to \eqref{ap} as
\begin{align}\label{go9}
	\hat{\mathbf{h}}(\mathbf{q}, \mathbf{u}) = \mathbf h(\mathbf q,\mathbf u)
	\bigl\lvert_{d=\bar d,\;\phi=\bar\phi,\;\theta=\bar\theta,\;\nu=\bar\nu}.
\end{align}

The proposed channel estimation method is summarized in Algorithm 1. Next, we analyze the complexity of the proposed channel estimation method by considering complex multiplications. The complexity of the surface-wise estimation and the multi-surface refinement estimation is \(\mathcal{O}(NB|\Xi_{\mathrm{H}}|_{\mathrm{c}}+B{N_c})\) and \(\mathcal{O}(SN|\Xi_{\mathrm{L}}|_{\mathrm{c}})\), respectively. The proposed channel estimation approach reduces the computational complexity by utilizing surface-based structures and incorporating only the reliable estimates from 6DMA position-rotation pairs.
	\section{Deep Learning-based 6DMA Position and Rotation Optimization}
	\subsection{Problem Formulation}
		\begin{figure*}[t!]
		\centering
		\setlength{\abovecaptionskip}{0.cm}
		\includegraphics[width=6.0in]{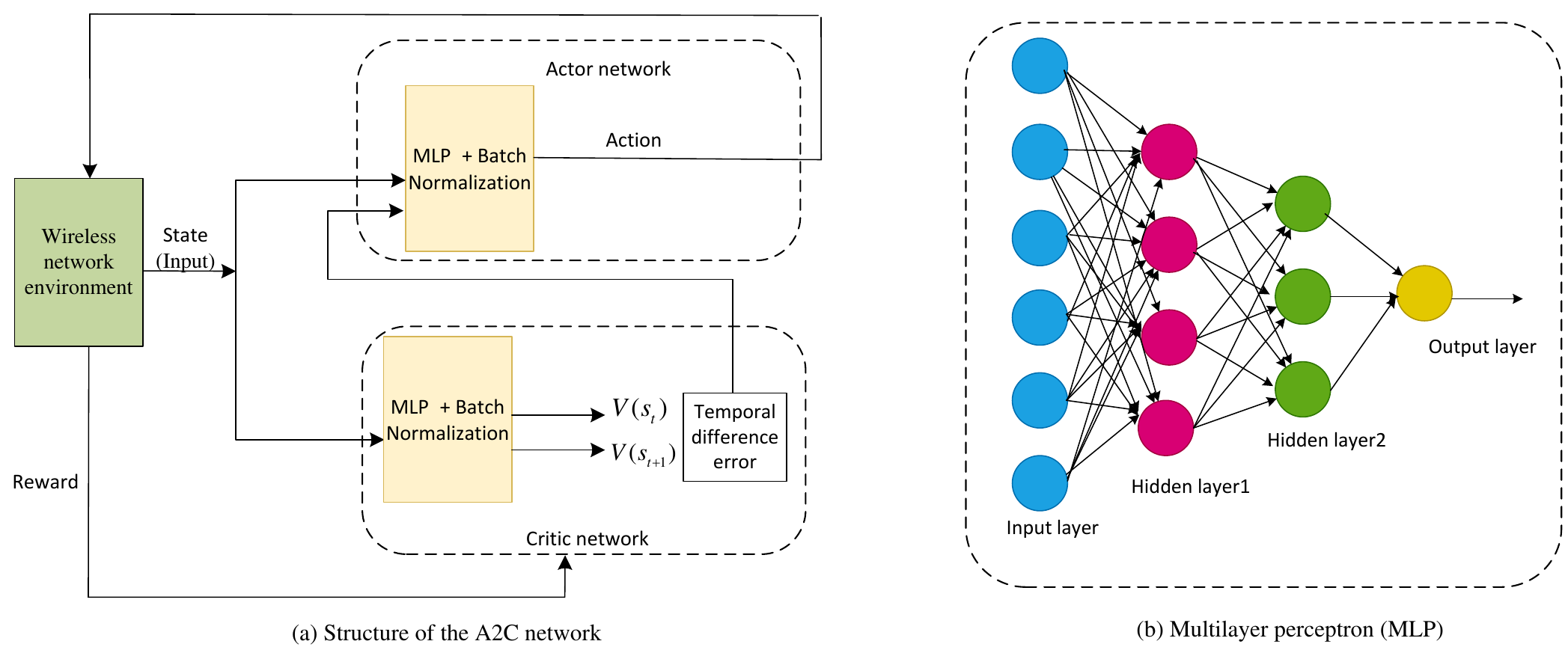}
		\caption{Deep reinforcement learning for joint optimization of 6DMAs' positions, rotations, and beamforming.}
		\label{DRL}
	\end{figure*}
	Based on \eqref{r} and the estimated channel, we aim to design the transmit beamforming, along with 6DMA surface's position and rotation to maximize the achievable sum-rate.
	For a given maximum
	transmit power budget $P_{\max}$, the design can be formulated as
	\begin{subequations}
		\label{MG3}
		\begin{align}
			\text{(P1)}~~&~\mathop{\max}\limits_{\{\mathbf{w}_k\}_{k=1}^K,\mathbf{q},
				\mathbf{u}}~~
			\sum_{k=1}^{K}C_k(\mathbf{w}_k,\mathbf{q},\mathbf{u})\\
			\text {s.t.}~&~\sum_{k=1}^{K}\|\mathbf{w}_k\|^2\leq P_{\max}, \label{M00}\\
			~&~\mathbf{q}_b\in\mathcal{C}, ~\forall b \in \mathcal{B}, \label{M1}\\
			~&~ \|\mathbf{q}_b-
			\mathbf{q}_{j}\|^2\geq d_{\min},~\forall b ,j \in \mathcal{B}, ~b\neq j,\label{M2}\\
			~&~ \mathbf{n}^T(\mathbf{u}_b)(\mathbf{q}_{j}-\mathbf{q}_b)\leq  0,~\forall b ,j \in \mathcal{B}, ~b\neq j, \label{M3}
		\end{align}
	\end{subequations}
	where \( \mathbf{n}(\mathbf{u}_b) = \mathbf{R}(\mathbf{u}_b)\bar{\mathbf{n}} \) with \( \bar{\mathbf{n}} \) being the normal vector of the \( b \)-th 6DMA surface in the local CCS.
	Constraint \eqref{M1} confines that the 6DMA center is located in its configuration convex region $\mathcal{C}$. The minimum distance $d_{\min}$ in constraint \eqref{M2} avoids any potential coupling effects between 6DMA surfaces, while constraint \eqref{M3} prevents mutual signal reflection between any two 6DMA surfaces \cite{shao20246d,6dma_dis}. Problem (P1) is non-convex, as the positions \(\mathbf{q}\) are coupled with rotations \(\mathbf{u}\) and beamforming \(\{\mathbf{w}_k\}_{k=1}^K\) in the objective function, which makes it challenging to obtain the globally optimal solution in general. To address this problem, we will design a low-complexity DRL-based algorithm, which aims to maximize the sum rate by observing rewards and iteratively adjusting its parameters accordingly.
	\subsection{Proposed A2C-based Deep Learning for 6DMA System}
	In practice, DRL is designed to enhance the ability to control and make decisions in a specific environment by employing a learning  method through experimentation and error correction \cite{DRL}. Let \(S\) represent a collection of all states. At each time-step \(t\), the agent receives a state \(s_t\in S\) and chooses an action \(a_t\) from the set of viable actions \(\mathcal{A}\) based on its policy \(\pi(a_t | s_t)\), which denotes the probability of taking action \(a_t\) conditioned on the state \(s_t\).
	Since A2C can achieve stable and fast learning outcomes \cite{a22c}, we develop a low-complexity A2C-based DRL approach to optimize beamforming and the position and rotation of 6DMAs. At each
	time step, the DRL agent chooses an action by interacting with
	the environment.  The structures of the A2C network adopted in this paper are illustrated in Fig. \ref{DRL}. We define the state space, action space, and reward function for the proposed DRL model as follows.
	\begin{figure}[t!]
		\centering
		\setlength{\abovecaptionskip}{0.cm}
		\includegraphics[width=3.2in]{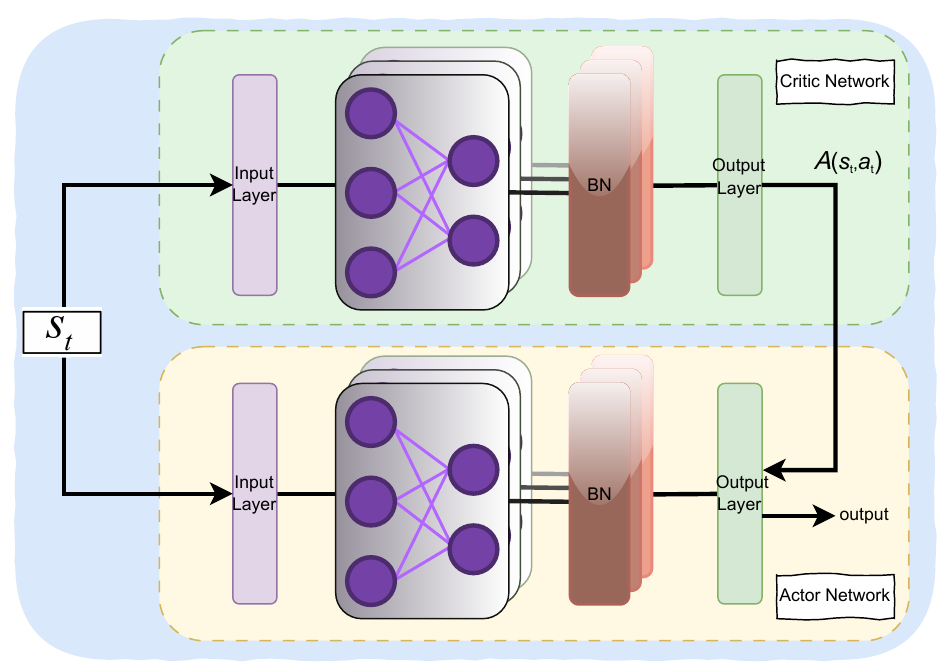}
		\caption{Proposed MLP structure of the critic network and the actor network.}
		\label{net}
		\vspace{-0.69cm}
	\end{figure}
	\subsubsection{Reward}
	The reward function is crucial in deep reinforcement learning problems and must be carefully designed to guide the agent toward the optimal policy. Indeed, reward functions are commonly related to
	the objective of the problem. Moreover, the DRL
	agent should be trained to learn the optimal parameters without violating the operation constraints in (P1). Therefore, at the $t$-th step of the DRL, we design the reward function as
	\begin{align}\label{pen}
		\!\!\!\!\!r\!=\!\left\{\begin{matrix}
			&\!\!\!\!\!\!\sum_{k=1}^{K}C_k(\mathbf{w}_k^{(t)},\mathbf{q}^{(t)},\mathbf{u}^{(t)}), \text{if system is normal},\\
			&\!\!\!\!\!\!-(P_{\mathrm{po}}\!+\!
			P_{\mathrm{dis}}\!+\!P_{\mathrm{ro1}}), \text{there are constraints violations},
		\end{matrix}\right.\!\!\!\!
	\end{align}
	where $P_{\mathrm{po}}$, $P_{\mathrm{dis}}$, and $P_{\mathrm{ro1}}$ correspond to the total amount of violations as the reward if any constraints violation. Specifically, as indicated by \eqref{M1}, for all 6DMA surfaces, if their positions are not within the allowed area $\mathcal{C}$, we impose the following penalty
	\begin{equation}\label{ppo}
		P_{\mathrm{po}} = \sum_{b \in B} \mathbb{I}(\mathbf{q}_b \notin \mathcal{C})\times \frac{L}{2},
	\end{equation}
	where $L$ represents the edge length of the cube $\mathcal{C}$ and $\mathbb{I}(\cdot)$ is an indicator function that returns 1 when the condition is true, otherwise it returns 0.
	Moreover, according to minimum distance constraint in \eqref{M2}, for each pair of 6DMA surfaces, if their distance is less than the minimum distance $d_{\min}$, we subtract the following penalty
	\begin{equation}\label{pds}
		P_{\mathrm{dis}} = \sum_{b \in B} \sum_{j \in B, j \neq b} 10{e^{\max(0, d_{\min} - \|\mathbf{q}_b - \mathbf{q}_j\|^2)}}.
	\end{equation}
	In addition, according to the rotation constraint in \eqref{M3}, for each pair of 6DMA surfaces, if the antennas exhibit mutual signal reflection, we impose the following penalty:
	\begin{equation}\label{ro1}
		P_{\mathrm{ro1}} = \sum_{b \in B} \sum_{j \in B, j \neq b} 10{e^{\max\left(0, \frac{\mathbf{n}^T(\mathbf{u}_b) (\mathbf{q}_j - \mathbf{q}_b)}{\|(\mathbf{q}_j - \mathbf{q}_b)\|}\right)}}.
	\end{equation}
	
	\subsubsection{State Space}
	In 6DMA systems, the state $s_t$ at time step $t$ comprises the transmission power, the received power of users, and the user locations $\mathbf{v}_k$ for $k=1,2,\cdots, K$. The transmission power for the $k$-th user is determined by the square of the magnitude of the transmit symbols, which have unit variance. Since neural networks process only real-valued numbers, any complex numbers involved in the state $s_t$ are split into their real and imaginary components. Mathematically, it is expressed as $\|\mathbf{w}_k\|^2 = |\Re\{\mathbf{w}_k^H\mathbf{w}_k\}|^2 + |\Im\{\mathbf{w}_k^H\mathbf{w}_k\}|^2$. The first term represents the real part's contribution, while the second term represents the imaginary part's contribution. Both components serve as separate inputs to the critic and actor networks. The state, $s_t$, resulting from the transmission power will have a total of $2K$ entries.
	We define the received power at the $k$-th user as $|(\mathbf{h}_k(\mathbf{q}, \mathbf{u}))^H \mathbf{w}_k|^2$, which can be further decomposed into $|\Re\{(\mathbf{h}_k(\mathbf{q}, \mathbf{u}))^H \mathbf{w}_k\}|^2 + |\Im\{(\mathbf{h}_k(\mathbf{q}, \mathbf{u}))^H \mathbf{w}_k\}|^2$. The total number of entries for the received power is $2K^2$. Additionally, the total number of entries for all user locations is $3K$. Therefore, the dimension of the state space is $5K + 2K^2$.
	
	In summary, the state space $s_t$ is defined as:
	\begin{align}\label{as}
		s_t=&\{\{|\Re\{\mathbf{w}_k^H\mathbf{w}_k\}|^2,|\Im\{\mathbf{w}_k^H\mathbf{w}_k\}|^2, |\Re\{\mathbf{h}_k^H \mathbf{w}_k\}|^2,\nonumber\\
		&|\Im\{\mathbf{h}_k^H \mathbf{w}_k\}|^2,\mathbf{v}_k \}_{k=1}^{K}\}.
	\end{align}
	
	\subsubsection{Action Space}
	In 6DMA systems, the action is defined by the transmit beamforming matrix $\{\mathbf{w}_k\}_{k=1}^K$, the position $\mathbf{q}$, and the rotation $\mathbf{u}$ of the 6DMAs. For a complex beamforming vector $\mathbf{w}_k = \Re\{\mathbf{w}_k\} + j\Im\{\mathbf{w}_k\}$, the real and imaginary parts are treated as separate entries. Thus, the action space $a_t$ is formulated as follows:
	\begin{align}\label{as}
		a_t=\{\mathbf{q}^{(t)},\mathbf{u}^{(t)}, \{\Re\{\mathbf{w}_k\}, \Im\{\mathbf{w}_k\}\}_{k=1}^K\}.
	\end{align}
	Note that the beamforming vectors $\mathbf{w}_k, \forall k\in\{1,2,\cdots,K\}$ should satisfy the power constraint given in \eqref{M00}. To ensure this, a normalization layer is applied at the output of the actor network to guarantee that $\sum_{k=1}^{K}\|\mathbf{w}_k^{(t)}\|^2= P_{\max}$.
	
	\subsubsection{Network Parameter Update}
	As illustrated in Fig. \ref{DRL}(a), the wireless network environment is modeled by the spatial distribution of user channels. The A2C-6DMA solution comprises two main components, namely the actor network and the critic network. The goal of the A2C network is to optimize the predicted reward obtained from each state $s_t$, denoted by $R_t=\sum_{i=0}^{\infty}\gamma^i r_{t+i}$, by utilizing the action-value function $Q(s_t, a_t)=\mathbb{E}[R_t|s_t=s, a_t=a]$ and the state-value function $V(s_t)=\mathbb{E}[\sum_{i=0}^{\infty}\tau^i r_{t+i}|s_t=s]$, which provides an estimation of the average expected return from state $s$ \cite{a22c}. Here, $\tau \in[0,1]$ is a discount factor.
	
	The actor network, which is responsible for action generation based on current states, is a multilayer perceptron (MLP) with two fully connected layers (see Fig. \ref{DRL}(b)). The first layer processes the state $s_t$ to extract features relevant to action selection, while the second layer uses a softmax function to convert its output into action probabilities $\pi(a_t, s_t)$. Actions are then sampled according to these probabilities.
	
	The critic network estimates state values and is also implemented as an MLP with two fully connected layers (see Fig. \ref{DRL}(b)). 	
	The first layer extracts value-related features from the state $s_t$, which are then processed by a second layer with a single neuron to output the state value $V(s_t)$. 
	
	In practice, the actor and critic networks are characterized by parameters \(z_a\) and \(z_c\), respectively. Next, we present the update expressions for these networks.
	In A2C network, the actor network's loss function is defined as \cite{a22c}
	\begin{align}\label{acc1}
		J(z_a)=\mathbb{E}[\ln \pi_{z_a}(a_t, s_t)A(s_t, a_t)],
	\end{align}
	where $A(s_t, a_t)=Q(s_t, a_t)-V(s_t)$ denotes advantage function, and the expectation
	reflects the empirical average across a small batch of data. Since the A2C aims to minimize the policy objective in \eqref{acc1}, we  calculate the gradient of $J(z_a)$ with respect to \( z_a \) \cite{a22c}, which is given by
	\begin{align}\label{deta}
		\nabla_{z_a} J(z_a)&=\mathbb{E}[\nabla_{z_a}\ln \pi_{z_a}(a_t, s_t)A(s_t, a_t)]\\
		&=\nabla_{z_a}\ln \pi_{z_a}(a_t, s_t)(Q(s_t, a_t)-V_{z_c}(s_t)),
	\end{align}
	where 
	$\nabla_{z_a}\ln \pi_{z_a}(a_t, s_t)$ is the gradient of the policy.
		\begin{algorithm}[t!]
		\caption{Deep Reinforcement Learning for Solving Problem (P1)}
		\label{alg2}
		\begin{algorithmic}[1]
			\STATE \textbf{Input}: Episodes size $I$, user locations $\mathbf{v}_k, \forall k$;  \\
			\STATE \textbf{Output}: Optimal action $a =\{\mathbf{q}, \mathbf{u}, \{\mathbf{w}_k\}_{k=1}^K\}$ and approximate optimal policy of the actor network $\pi$; \\
			\STATE \textbf{Initialization}: Initialize 6DMAs' position $\mathbf{q}$, 6DMAs' rotation $\mathbf{u}$, transmit beamforming $\{\mathbf{w}_k\}_{k=1}^K$, and actor-critic networks $z_a$ and $z_c$;
			\FOR {$episode=1,2,\cdots,I$}
			\STATE Collect and preprocess transmission power, received
			power of users, and the user locations for the $i$-th episode
			to obtain the first state $s_1$;
			\FOR {$t=1,2,\cdots,T$}
			\STATE The actor network processes the input \( s_t \) and computes the action probability distribution \( \pi(a_t | s_t) \);
			\STATE The critic network processes the input \( s_t \) and computes the state-value \( V(s_t) \);
			\STATE The agent selects an action \( a_t \) based on the probability distribution \( \pi(a_t | s_t) \) and executes \( a_t \);
			\STATE The agent receives a reward \( r_t \) and observes the new state \( s_{t+1} \);
			\STATE The critic network processes the new state \( s_{t+1} \) and computes its state-value \( V(s_{t+1}) \);
			\STATE Compute the temporal difference error \( \delta(s_t) = r_t + \tau V(s_{t+1}) - V(s_t) \);
			\STATE Update the critic network parameters \( z_c \) according to \eqref{lossa};
			\STATE Update the actor network parameters \( z_a \) according to \eqref{deta3};
			\ENDFOR
			\ENDFOR
		\end{algorithmic}
	\end{algorithm}
	
	Thus, the actor network is updated leveraging
	\begin{align}\label{detaa1}
		z_a=z_a+\mu_a\nabla_{z_a}\ln \pi_{z_a}(a_t, s_t)(Q(s_t, a_t)-V_{z_c}(s_t)),
	\end{align}
		where $0<\mu_a\leq 1$ denotes the learning rate for actor network.
Given that the temporal difference (TD) error, $\delta_{z_c}(s_t)=r_t+\tau V_{z_c}(s_{t+1}|s_t, a_t)-V_{z_c}(s_{t})$, introduced in \cite{a22c} provides an unbiased estimate of the advantage function, it can replace \(Q(s_t, a_t) - V_{z_c}(s_t)\) in the update equation \eqref{detaa1}. This yields the update
	\begin{align}\label{deta3}
		z_a=z_a+\mu_a\nabla_{z_a}\ln \pi_{z_a}(a_t, s_t)\delta_{z_c}(s_t).
	\end{align}

	On the other hand, the critic network is updated iteratively using the mean square error (MSE) loss function, which is defined as
	\begin{align}\label{loss}
		\tilde{J}=\delta_{z_c}^2(s_t).
	\end{align}
	Accordingly, the critic network parameters can be updated according to \cite{a22c}, i.e.,
	\begin{align}\label{lossa}
		z_c=z_c+\mu_c\delta_{z_c}(s_t)\frac{\partial V_{z_c}(s_t)}{\partial z_c},
	\end{align}
	where $0<\mu_c\leq 1$ denotes the learning rate for critic network.
	
	One challenge in training the proposed A2C network is the non-stationary distribution of each layer's inputs during training, which requires the use of lower learning rates and careful parameter initialization. As shown in Fig. \ref{net}, this problem is addressed by applying batch normalization (BN) \cite{cobbe2019quantifying} to the inputs of each MLP hidden layer for each training mini-batch. This allows for higher learning rates and less concern with initialization, thereby improving the model's generalization and accuracy.
	\vspace{-3 mm}
	\subsection{Initialization and Complexity Analysis}
	\vspace{-1 mm}
	For initialization, we propose generating $\varpi > B$ candidate positions uniformly on a spherical surface with the maximum feasible radius within BS site space $\mathcal{C}$. We assume that each candidate position is associated with a single rotation, as defined in \cite{6dma_dis}. From the generated candidate set $\{1, 2, \ldots, \varpi\}$, the initial positions and rotations are randomly chosen. 
	
	We outline detailed pseudocode  of the proposed DRL algorithm's in Algorithm 1. Over multiple episodes, the algorithm performs a number of training steps. In each episode, an action is taken and a reward is returned. Then, using the advantage function, the neural network weights are updated accordingly. The proposed algorithm comprises two main parts: the learning process and the fully connected layer model. 
	First, the neural network includes a fully connected model with two hidden layers, with complexity \(\mathcal{O}(\sum_{l=1}^{A} Z_l Z_{l-1})\), where \(Z_l\) represents the number of neurons in layer \(l\). Second, the learning process has a complexity of \(\mathcal{O}(|S|_{\mathrm{c}} |\mathcal{A}|_{\mathrm{c}})\), where \(|S|_{\mathrm{c}}\) and \(|\mathcal{A}|_{\mathrm{c}}\) denote the total number of states and actions, respectively. 
%	Overall, the computational complexity of the DRL-based algorithm scales linearly with the number of states and actions and quadratically with the number of neurons in each layer.
	
	\section{Simulation Results}
		\begin{table}[!t]
		\small
		\caption{{Simulation Parameters.}}
		\label{Table1}
		\centering
		\begin{tabular}{|c|c|c|c|c|}
			\hline
			Symbol & Description &Value \\
			\hline
			$I$&\makecell[c]{Number of episodes}&5,000 \\
			\hline
			$T$&\makecell[c]{Number of steps\\ in each episode}&20,000  \\
			\hline
			$N$&\makecell[c]{Number of antennas of \\each 6DMA surface}&16  \\
			\hline
			$B$&Number of 6DMA surfaces & 8 \\
			\hline
			$\mathcal{C}$& 6DMA-enabled BS site space & Cube with 2 m sides\\
			\hline
			$d$&  User-BS distance & 20-250 m\\
			\hline
			$\sigma^2$& Average noise power & -80 dBm \\
			\hline
			$\lambda$ &Carrier wavelength & 0.125 m \\
			\hline
			$W$& Number of hotspots & 3\\
			\hline
			$\bar{d}$& \makecell[c]{Minimum antenna spacing\\
				on 6DMA surface}
			& $\lambda/2$\\
			\hline
			$\varpi$& \makecell[c]{Number of positions \\for initialization using \\Fibonacci Sphere}
			& 64\\
			\hline
		\end{tabular}
	\end{table}
	\begin{figure}[t!]
		\centering
		\setlength{\abovecaptionskip}{-3pt}
		\setlength{\belowcaptionskip}{-3pt}
		\includegraphics[width=3.3in]{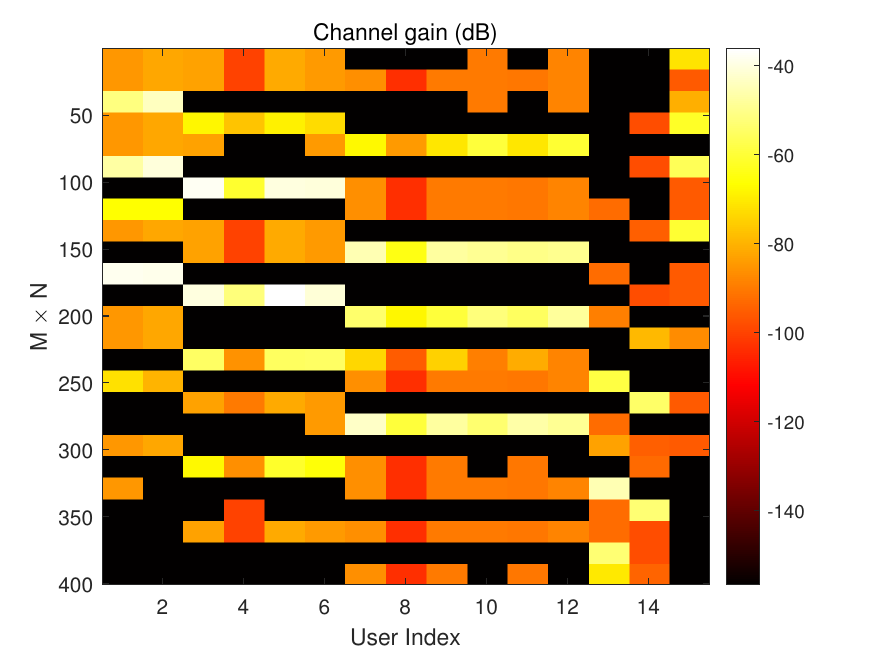}
		\caption{An illustration of the directional sparsity in the LoS 6DMA channel.}
		\label{heatmap}
		\vspace{-0.49cm}
	\end{figure}
	
In the simulation, the region \(\mathcal{C}\) forms a cube with side length \(L=2\) m. The array aperture of the 6DMA BS is given by \(D=L\sqrt{3}\) m. Applying the Rayleigh distance formula \(\mathrm{RD}=2D^{2}/\lambda\) \cite{nearfar}, the near-field-far-field boundary in this setup equals 192 m.
	%All simulation results are averaged over 500 independent channel realizations.
	Users are situated within a 3D coverage area denoted as \(\mathcal{L}\), which forms a spherical annulus ranging from $20$ m to $200$ m in radial distance from the BS center. The area \(\mathcal{L} = \mathcal{L}_0 \cup (\cup_{w=1}^3 \mathcal{L}_w)\) is segmented into four distinct subareas: three non-overlapping hotspot regions, \(\mathcal{L}_1\), \(\mathcal{L}_2\), and \(\mathcal{L}_3\), and the remaining region, \(\mathcal{L}_0\). Each hotspot area is shown as a 3D spherical region. Three hotspot regions are centered at distances of $40$ m, $60$ m, and $100$ m, respectively, from the reference position of the 6DMA-enabled BS. The radii of these spherical areas are set to $5$ m, $10$ m, and $15$ m, respectively.
	The spatial distribution of users within each hotspot area \(\mathcal{L}_w\) for $w\in\{1,2,3\}$ is characterized as an independent finite homogeneous Poisson point process (HPPP) with a density parameter \(\mu_w\) \cite{hppp}. The densities \(\mu_w\) are chosen such that the average number of users in \(\mathcal{L}_1\), \(\mathcal{L}_2\), and \(\mathcal{L}_3\) follows a ratio of 1:2:3.
	Additionally, within the entire coverage area \(\mathcal{L}\), the proportion of non-hotspot users among the total number of users is set to 0.2.
	The minimum distance \(d_{\min}\) between antennas on the UPA antenna surface is defined as \(\frac{\sqrt{2}}{2}\lambda + \frac{\lambda}{2}\). The effective antenna gain for the 6DMA-enabled BS  follows the 3GPP standard \cite{3gpp, shao20246d}. Table II lists the key simulation parameters. The learning rates for training actor-network and critic network are set as follows
	\begin{align}\label{pen}
		\mu_a=\mu_c=\left\{\begin{matrix}
			&0.1,~ \text{if} ~r<0,\\
			&0.01, ~\text{if} ~r=0,\\
			&10^{-5},~ \text{if} ~r>0.
		\end{matrix}\right.
	\end{align}
\begin{figure}[t!]
		%\vspace{-0.49cm}
	\centering
	\setlength{\abovecaptionskip}{-3pt}
	\setlength{\belowcaptionskip}{-3pt}
	\includegraphics[width=3.3in]{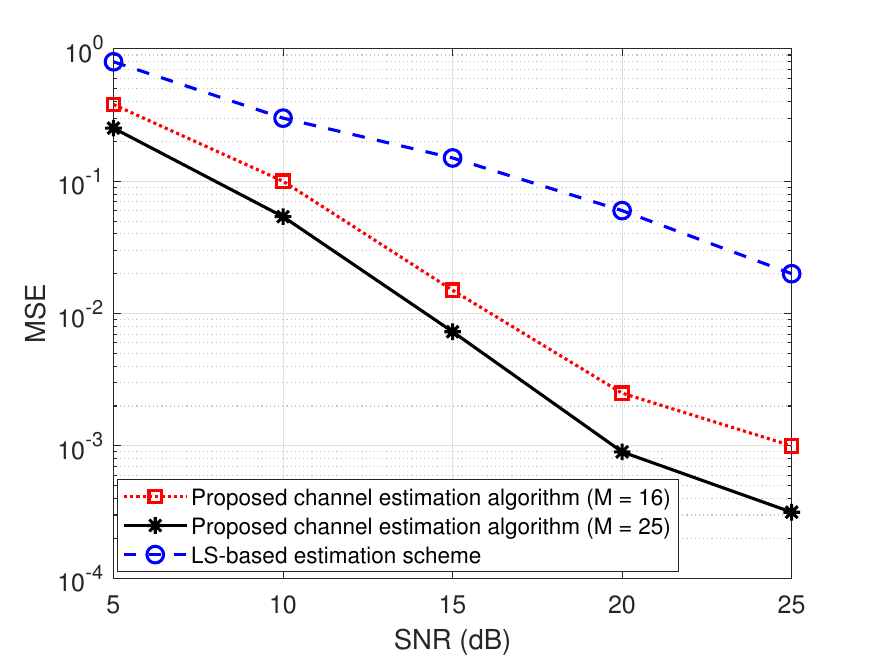}
	\caption{The MSE of 6DMA  channel estimation versus SNR.}
	\label{snrc}
	\vspace{-0.49cm}
\end{figure}

For comparison, we evaluate several benchmark schemes using a three-sector BS configuration, which is viewed as a specific 6DMA setup with \( B = 3 \) sectors and around \(\lceil\frac{NB}{3}\rceil\) antennas per surface, each covering roughly \(120^\circ\). The schemes include: (1) FPA; (2) FAS, where the rotation of all antennas is fixed and the proposed DRL-based algorithm is employed to optimize the position of each antenna within its respective sector; and (3) rotatable 6DMA, where the center position of each sector is fixed and the proposed DRL-based algorithm optimizes each sector antenna’s rotation.
	
For performance comparison in channel estimation, we assume that \( M \) candidate position-rotation pairs used for training measurements are evenly distributed across the largest spherical surface that fits within the 6DMA-BS site space \(\mathcal{C}\). 
Moreover, we consider the least squares (LS)-based estimation scheme as the baseline. In this scheme, for a given user, signals from $M$ 6DMA position-rotation pairs are jointly processed by concatenating them into a single observation vector. An LS-based channel estimate is then obtained from this combined data, and a grid search over user distance and angles is performed based on the channel estimate to extract the user’s location parameters. 

First, Fig. \ref{heatmap} illustrates the directional sparsity of the 6DMA channels from the \( K \) users to the \( M \) measured 6DMA position rotation pairs. It is observed that users are served by only a limited number of 6DMA position rotation pairs. In addition, the block sparsity indicates that a 6DMA surface is the smallest unit that exhibits stationarity, that is, an equal channel power distribution across antennas on the same 6DMA surface.

To evaluate the channel reconstruction performance, Fig. \ref{snrc} shows the mean MSE of channel estimation versus the received signal-to-noise ratio (SNR) for the proposed channel estimation algorithm and LS-based scheme. We observe that the proposed algorithm yields a significantly lower MSE compared to the LS-based method. This is because the proposed approach uses clustering to remove outliers and highly erroneous estimates caused by directional sparsity and retains only the most representative cluster for more accurate user localization. The LS‑based scheme does not exploit this clustering step. Moreover, increasing the number of candidate position-rotation pairs further improves channel estimation accuracy, since more 6DMA surfaces capture the user’s signals from different directions.
		\begin{figure}[t!]
		\centering
		\setlength{\abovecaptionskip}{0.cm}
		\includegraphics[width=3.3in]{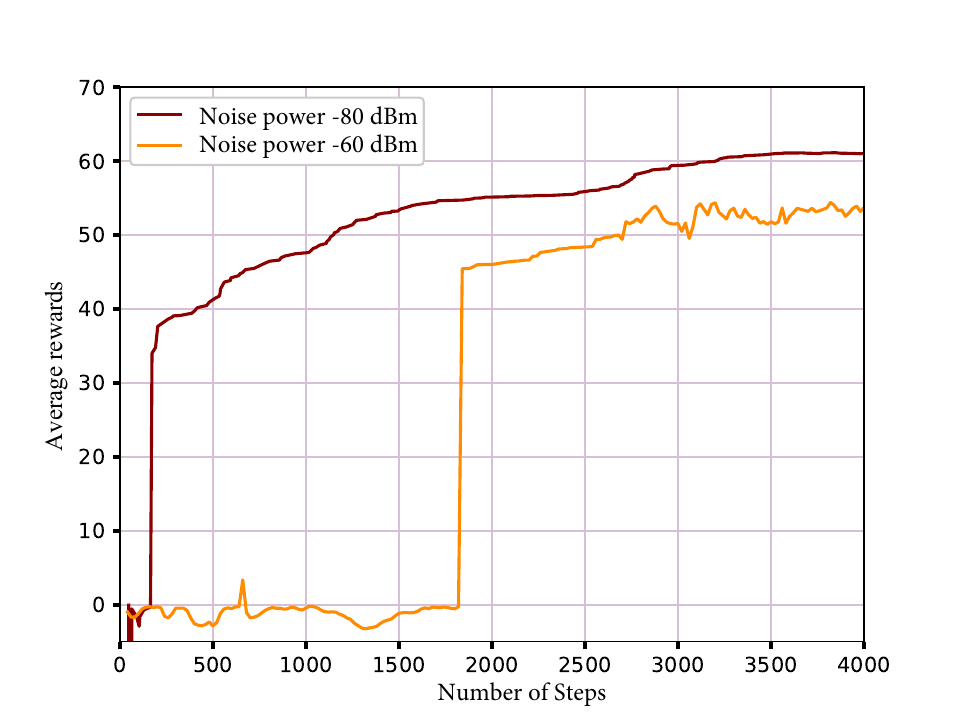}
		\caption{Rewards versus time steps under different noise powers.}
		\label{Power}
				\vspace{-0.49cm}
	\end{figure}
	
	\begin{figure}[t!]
		\centering
		\setlength{\abovecaptionskip}{0.cm}
		\includegraphics[width=3.3in]{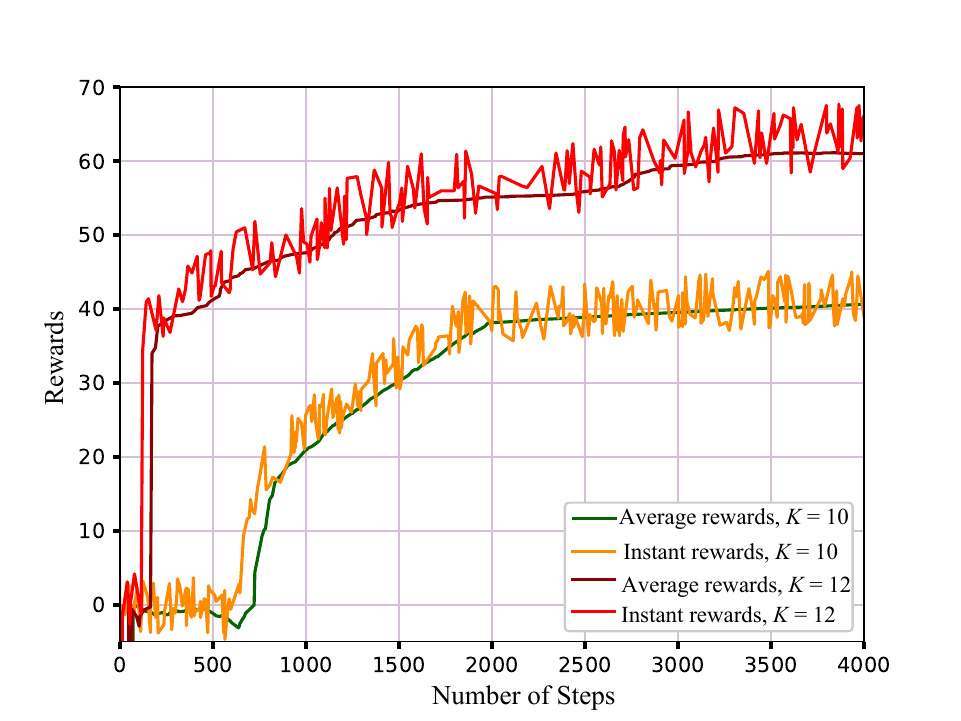}
		\caption{Rewards versus time steps under different number of users.}
		\label{Instant}
					\vspace{-0.49cm}
	\end{figure}
	
	To better understand the proposed DRL-based method, we analyze the impact of noise power and the number of users on its performance using the hybrid-field 6DMA channel model, as illustrated in Fig. \ref{Power} and Fig. \ref{Instant}, respectively. In Fig. \ref{Instant}, we examine both instant rewards and average rewards over time steps. In our simulations, we employ the following approach to compute the average rewards
	\begin{align}\label{av}
		\text{average reward}=\frac{\sum_{m=1}^{J_i} r_m}{J_i}, J_i=1,2,\cdots,J,
	\end{align}
	where $J$ is the maximum steps. From these two figures, it is evident that the rewards converge as the number of time steps increases. This indicates that the DRL-based algorithm starts from initial points and learns from the environment, gradually adjusting $\mathbf{q}$, $\mathbf{u}$, and $\{\mathbf{w}_k\}_{k=1}^K$ towards optimal solutions. In Fig. \ref{Power}, it is apparent that higher noise powers significantly affect the convergence rate and overall performance, particularly in scenarios with higher noise levels. Furthermore, from Fig. \ref{Instant}, it is observed that 6DMA with the proposed DRL converges faster with fewer users compared to more users. The reason is that, with a larger number of users, the dynamic range of instant rewards is larger, leading to more fluctuations and poorer convergence.	
   \begin{figure}[t!]
   	%	\vspace{-0.49cm}
	  \centering
	  \setlength{\abovecaptionskip}{0.cm}
	  \includegraphics[width=3.2in]{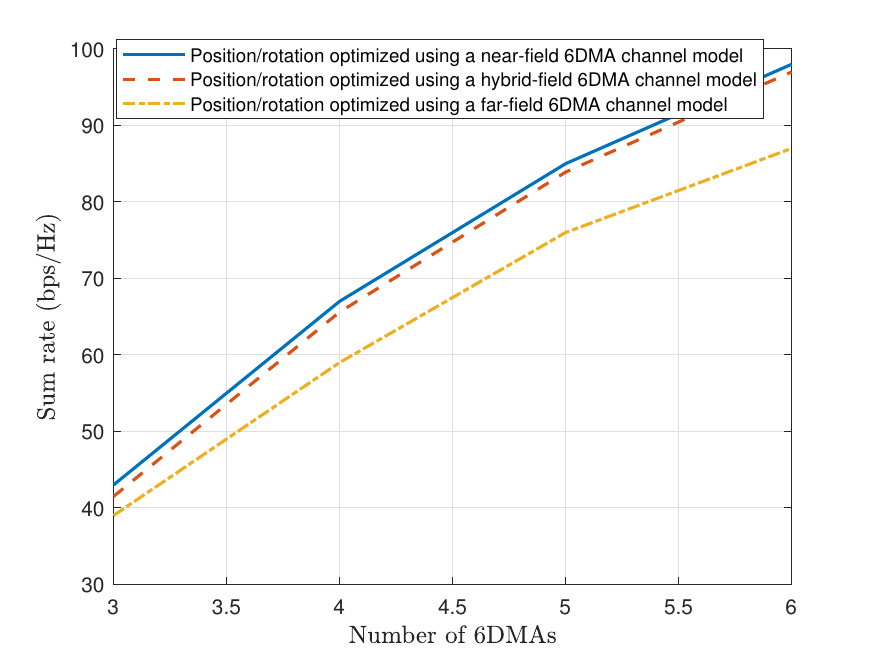}
	  \caption{Sum rate versus the number of 6DMA surfaces in the near-field 6DMA channel models.}
	  \label{nearfar}
	  		\vspace{-0.49cm}
	\end{figure}
	
In Fig. \ref{nearfar}, we first apply the proposed DRL method to the near-field, hybrid-field, and far-field 6DMA channel models to determine the corresponding sets of antenna positions, rotations, and beamformers. Subsequently, we calculate the sum rate for the near-field channel model using each set of antenna positions, rotations, and beamformers, and compare the performance of the three schemes. It is observed that the average sum rate of the hybrid-field 6DMA channel model closely approaches that of the near-field 6DMA channel model, which is significantly higher than that of the far-field 6DMA channel model.
This is because the far-field model ignores spherical-wave effects and fails to accurately capture channel characteristics in the near-field transmission. In contrast, the hybrid-field 6DMA channel model retains spatial multiplexing capability even with a single propagation path. Moreover, the hybrid-field 6DMA channel model utilizes fewer channel parameters than the near-field model while maintaining high estimation accuracy. As a result, the hybrid-field 6DMA channel model achieves near-optimal performance with reduced complexity.

	\begin{figure}[t!]
		\centering
		\setlength{\abovecaptionskip}{0.cm}
		\includegraphics[width=3.3in]{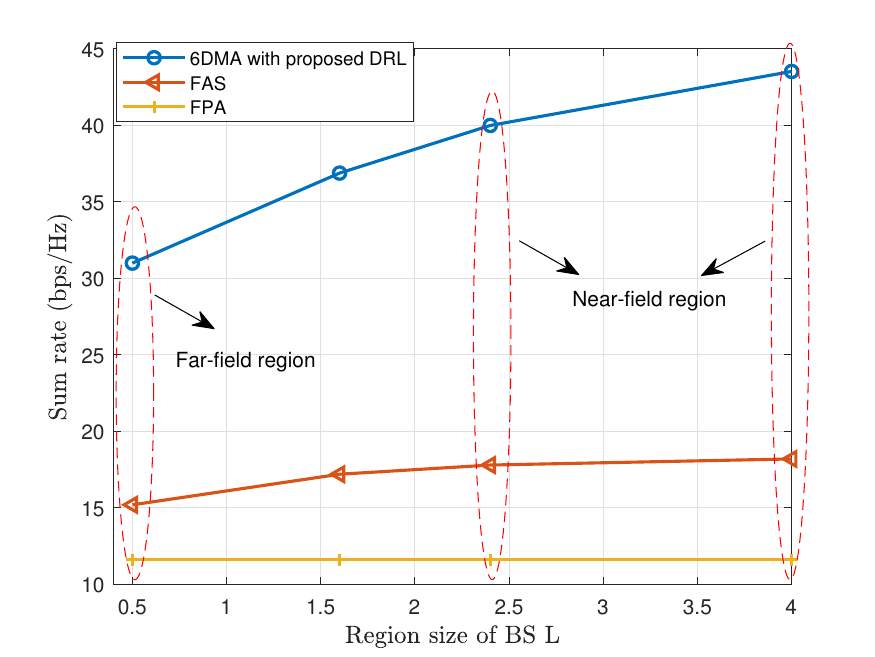}
		\caption{Sum rate versus region size of BS with $K=10$.}
		\label{regionsize}
				\vspace{-0.49cm}
	\end{figure}
	
In the subsequent simulations, the proposed DRL method is first applied using the hybrid-field 6DMA channel model to determine optimal antenna positions, rotations, and beamformers. These optimized parameters are then used to compute the sum rate under the near-field channel model, and the performance is compared against that of FPA and FAS schemes. In Fig. \ref{regionsize}, we show the sum rate versus the BS region size for the 6DMA with proposed DRL and the benchmark schemes. 
For \(L=0.5\) m, all users lie in the far field of every 6DMA surface. As \(L\) grows, they enter the near field. In every case, the 6DMA with the proposed DRL achieves a higher sum rate than FPA and FAS by exploiting extra spatial DoFs to match the channel’s spatial distribution. The advantage widens in the near field, which underscores its suitability for extremely large-scale MIMO. Furthermore, the FAS sum rate plateaus once the BS region exceeds 1.6 m, whereas the 6DMA with DRL continues to climb, showing no throughput limit from BS size.

%We know that when $L=0.5$, all users are located in the far-field region with respect to different 6DMA surfaces. As the region size of the BS continues to enlarge, the users are located in the near-field. It is observed that the 6DMA with proposed DRL outperforms FPA and FAS in terms of sum rate. This is because 6DMA-enabled BS is endowed with more spatial DoFs to adapt to the spatial distribution of the wireless channel.
%	Furthermore, the performance gain of the proposed 6DMA is more significant in the near-field region than in the far-field region. This indicates that the proposed 6DMA with DRL algorithm is appealing for the extremely large-scale MIMO in the near field. Moreover, the sum rate of the FAS scheme saturates when the BS region size is larger than $1.6$ m. However, the performance of the 6DMA with proposed DRL continues to improve as the BS region size increases. This indicates that the maximum sum rate of the 6DMA with the proposed DRL is not limited by the BS region size and does not reach a performance bottleneck.
		\begin{figure}[t!]
		%	\vspace{-0.49cm}
		\centering
		\setlength{\abovecaptionskip}{0.cm}
		\includegraphics[width=3.3in]{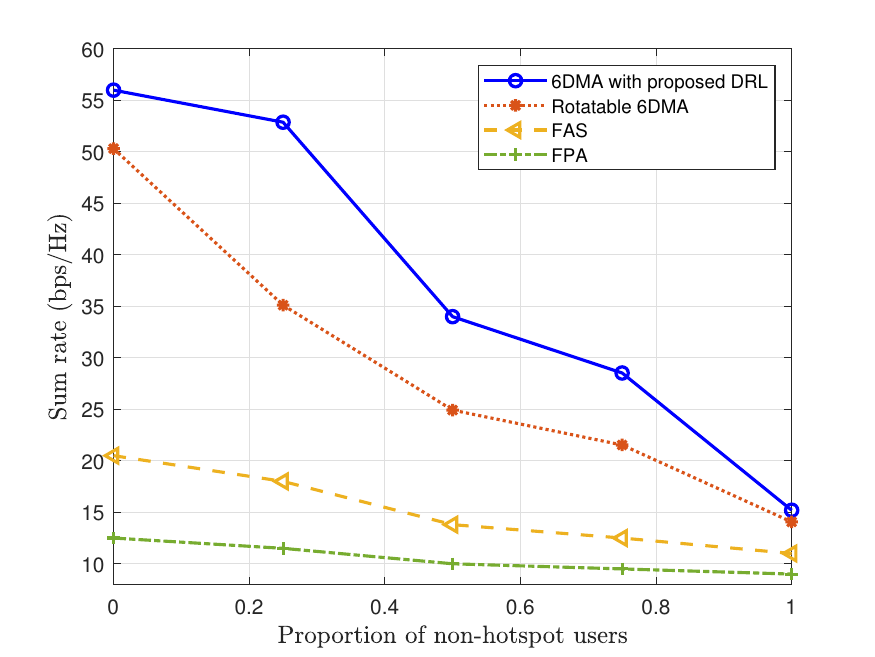}
		\caption{Sum rate versus the proportion of regular users.}
		\label{regular}
				\vspace{-0.49cm}
	\end{figure}
	
	Fig. \ref{regular} illustrates the impact of the proportion of non-hotspot users among the total number of users on the sum rate.  We see that the performance of all the algorithms considered decreases as the proportion of non-hotspot users increases. Additionally, the proposed algorithm consistently outperforms FPA and FAS across all proportions of non-hotspot users, although the performance gain diminishes as the proportion of non-hotspot users approaches one. These results indicate that the 6DMA with the proposed DRL is particularly advantageous when the spatial distribution of users is more non-uniform.
	
\begin{figure}[t!]
		\centering
		\setlength{\abovecaptionskip}{0.cm}
		\includegraphics[width=3.3in]{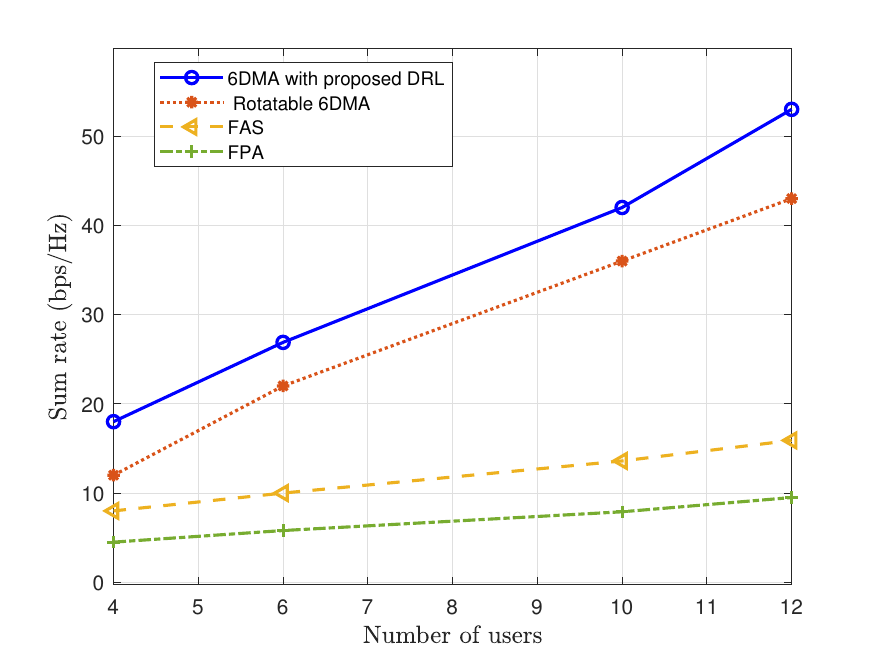}
		\caption{Sum rate versus the number of users.}
		\label{users}
						\vspace{-0.49cm}
	\end{figure}

Finally, Fig. \ref{users} compares the sum-rate performance of the proposed method and benchmark schemes as the user number increases. Although all schemes benefit from more users, the proposed DRL-enhanced 6DMA shows an expanding performance advantage. This is due to the improvement in spatial multiplexing gain of 6DMA with the proposed DRL with a higher number of users. These observations indicate that the 6DMA with the proposed DRL is particularly advantageous in more heavily loaded networks.
	
\section{Conclusion}
In this work, we introduced a novel hybrid-field 6DMA channel model to accurately model the 6DMA channel with a limited number of channel characteristics. We then proposed a low-complexity channel estimation algorithm by taking advantage of the non-stationary premise that not all users are served by all 6DMA position-rotation pairs. 
Subsequently, we developed a DRL-based optimization algorithm to overcome the high-dimensional complexity problem in the joint design of transmit beamforming, 6DMA positions, and rotations. The proposed DRL algorithm efficiently learns from the environment with low computational complexity and offers efficient implementation without relying on explicit wireless environment models or specific mathematical formulations. Simulation results validated the accuracy of the hybrid-field 6DMA channel model and confirmed the efficiency of the proposed channel estimation and DRL-based position/rotation optimization algorithms.
 	%\vspace{-0.29cm}
	\bibliographystyle{IEEEtran}
	\bibliography{fabs}
\end{document}